\documentclass{ws-ijmpe}
\def\draftnote{~}%
\def\trimmarks{}

\def\A{{\bm A}}
\def\B{{\bm B}}
\def\D{{\bm D}}
\def\E{{\bm E}}
\def\F{{\bm F}}
\def\k{{\bm k}}
\def\p{{\bm p}}
\def\v{{\bm v}}
\def\x{{\bm x}}
\def\y{{\bm y}}
\def\kB{k_{\rm B}}
\def\bnum{{\bm\#}}
\def\mD{m_{\rm D}}
\def\xiD{\xi_{\rm D}}
\def\grad{{\bm\nabla}}

\begin{document}

\markboth{Peter Arnold}{Quark-Gluon Plasmas and Thermalization}

\title{QUARK-GLUON PLASMAS AND THERMALIZATION
\protect\footnote{
  Based on lectures given at X Hadron Physics in Florian\'opolis, SC, Brazil,
  March 26--31, 2007.
}
}

\author{\footnotesize PETER ARNOLD}

\address{Department of Physics, University of Virginia, P.O.\ Box 400714\\
         Charlottesville, VA 22904-4714, U.S.A.\\
         parnold@virginia.edu}

\maketitle


\begin{abstract}
In these lectures, I will attempt a pedagogical and qualitative
introduction to the theory of equilibrium and thermalization of
quark-gluon plasmas.  I assume only that the reader is familiar with
quantum field theory at zero temperature and with QCD as the theory of
the strong interactions.  I focus on the limit of small
$\alpha_{\rm s}$, which in principle should be relevant at extremely
high temperature because of asymptotic freedom, and in any case
provides a clean theoretical context in which to discuss a variety of
phenomena.
Topics discussed include the basic equilibrium formalism for
finite-temperature quantum field theory, Debye screening,
electric deconfinement,
magnetic confinement,
dimensional reduction,
plasma waves, kinetic theory,
hydrodynamic properties
such as viscosity,
the Landau-Pomeranchuk-Migdal effect,
thermalization in (arbitrarily
high energy) heavy ion collisions, and QCD plasma instabilities.
\end {abstract}


\section{Introduction to Quark-Gluon Plasma}

Imagine a perfect container filled with nothing but vacuum.  What
happens if we slowly heat it?  At first, blackbody photons appear inside
the container.
As we increase the temperature, the density (and energy) of blackbody
photons increases.
Once we raise the temperature to the neighborhood of
$m_{\rm e} c^2 \simeq 0.5$ MeV,
something interesting happens.
The photons have enough energy to pair produce and make pairs of
electrons and positrons.

Now consider even higher temperature of order $\kB T \sim 50$ MeV.  A
few photons will have enough energy to pair produce pions ($m_\pi c^2
\simeq 135$ MeV), yielding a dilute gas of pions inside the container
(in addition to the photons and electrons).  Each individual pion, of
course, is a finite-size object made up of a quark and anti-quark in the
non-relativistic quark model (and containing other quark/anti-quark
pairs and gluons if we look close enough to resolve the structure on
smaller distance scales), held together by confinement.  As we continue
to increase the temperature, the density of pions increases and
increases, until eventually the pions are so dense that they overlap at
temperatures of order the QCD scale, $\kB T \sim 200$ MeV.  And if the
pions overlap, how is a quark to know which other partons it is bound to
by confinement?  This is a standard cartoon picture of deconfinement at
high temperature.  At high enough temperature, the quarks and gluons
become so dense that they lose track of who they are confined with and
so can individually wander across the system.  The important point
here is that, for ultra-relativistic temperatures, higher temperature
means higher {\it density}\/ as well as higher energy.

As we increase the temperature further to $\kB T \gg 200$ MeV, the
density of quarks and gluons increases, just as the density of photons
did.  And since their energies are of order $T$, one expects that the
running coupling $\alpha_{\rm s}$ should be $\alpha_{\rm s}(T)$.
By asymptotic freedom, this gas of quarks and gluons will become
weakly coupled if we go to very large $T$.

Historically, the temperatures necessary for a quark-gluon plasma (QGP)
existed up until the Universe was about a millionth of a second old.
Experimentally, quark gluon plasmas are believed to be created in
sufficiently energetic heavy ion collisions, such as the RHIC experiment
at Brookhaven National Lab, where, for example, gold nuclei are
collided at energies of order 100 GeV per nucleon.

In the context of heavy ion collisions,
this lecture will nicely complement the lecture by Francois Gelis\cite{gelis}
in this series.  In a heavy ion collision at very high energies, the
theoretical picture is that the colliding ions create
a ``color glass condensate,'' which eventually thermalizes to make
a quark-gluon plasma in local thermal equilibrium.  Francois's lectures
take us forward in time up through the creation of the color glass
condensate.  I will instead work backwards in time, starting from
a description of equilibrium.  Then I discuss the relaxation of
small departures from equilibrium (relevant to slightly
earlier times).  I conclude with the physics of
equilibration from very non-equilibrium initial conditions such
as the color glass condensate.


\subsection{Introduction to weak coupling}

A theorist's first instinct when confronted with any field theory
problem is generally to wonder if it can be solved for weak coupling.
In this case, that means $\alpha_{\rm s}(T) \ll 1$.

Let me first give you some examples of static equilibrium quantities.
The free energy has the weak coupling expansion\cite{F}
\begin {equation}
  F = \# T^4 [ 1 + \# g^2 + \# g^3 + g^4(\# \ln g + \#)
          + \# g^5 + g^6(\# \ln g + \bnum) + \cdots ,
\end {equation}
where all the numerical coefficients $\#$ are known except for the last
(boldface) one.  It may seem a little strange that the expansion has odd
powers of $g$ and factors of $\ln g$, since perturbation theory is
generally an expansion in $g^2$.  What's stranger is that the last,
unknown numerical coefficient (the coefficient of $g^6$) is not a
perturbative quantity at all: its value depends on fundamentally
non-perturbative physics (as do yet other coefficients at yet higher
orders).  We'll discuss how this can be in a moment, but it exemplifies
an important point for studying thermal gauge theories at weak
coupling: the weak coupling expansion of a physical quantity is
not necessarily the same things as the {\it perturbative}\/ or loop
expansion of that quantity that one learns in field theory courses.

An example where non-perturbative physics crops up sooner in the
expansion is the inverse Debye screening length,\cite{debye}
\begin {equation}
  \xi_{\rm D}^{-1} \simeq \#gT[1 + g(\# \ln g + \bnum)],
\end {equation}
where the last numerical coefficient turns out to be non-perturbative.
I will discuss leading-order Debye screening a little later.

Here is an example from real-time near-equilibrium physics.
The result for the shear viscosity of a QGP to leading order
in powers of the coupling is known\cite{shear}
and is of
the form
\begin {equation}
  \eta = \frac{\# T^3}{g^4 \ln g} \times f\left( \frac{1}{\ln g} \right)
  .
\end {equation}
We'll see that this is immensely complicated in terms of diagrammatic
perturbation theory.  Another example (which I won't pursue in these
lectures) comes from electroweak, rather
than QCD plasmas, where the rate of baryon number violation in the
very early universe is known to have the form\cite{B}
\begin {equation}
   \Gamma_{\rm B} \simeq {\bnum g^{10} T^4}{\ln g} \,,
\end {equation}
where the numerical coefficient is non-perturbative.

Now turn to real-time, far-from-equilibrium physics.
One question to ask is how long it takes for the system to
equilibrate.  Let's use dimensional analysis to write the
leading-order answer to this question in the form
\begin {equation}
   t_{\rm eq} \sim \frac{g^{-??}}{\mbox{momentum scale}} ,
\end {equation}
where ``momentum scale'' is the relevant momentum scale of the
problem.  The situation here is so bad, that not even the {\it power}\/
of $g$ is known in the weak coupling limit!


\subsubsection{Small coupling $\not=$ perturbative}
\label {sec:QMexample}

\begin{figure}[th]
\centerline{\psfig{file=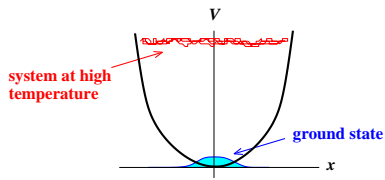,width=5cm}}
\caption{%
  A slightly anharmonic potential.
  \label {fig:anharmonic}
}
\end{figure}

So why is the small coupling expansion not the same as the
perturbative expansion at finite temperature?  There's a really
simple example from ordinary quantum mechanics that helps make
it clear.  Imagine a particle moving in a slightly anharmonic potential
\begin {equation}
  V(x) \sim \omega_0^2 x^2 + g^2 x^4 ,
\end {equation}
where $g$ is very, very small.  If we ask questions about the ground
state, say, then we can approximate the potential by a harmonic
oscillator and treat the quartic piece as a perturbation.
The ground state wave function will extend over a range of $x$'s
where $g^2 x^4$ is small compared to $\omega_0^2 x^2$.  But
now imagine studying the system at high temperature, weighting the
states by $\exp(-\beta E)$.  As the temperature is increased, the
typical energy $E$ will increase.  But high energy states
probe a large range of $x$.  (See Fig.\ \ref{fig:anharmonic}.)
The higher the energy, the larger
the typical values of $x$.  No matter how small $g$ is, $g^2 x^4$
will always be bigger than $\omega_0^2 x^2$ for large enough $x$.
So, at high enough temperature, a perturbative
treatment of $g^2 x^4$ breaks down.

Note that in this example the break down of perturbation theory
corresponds to high energies, which is also the limit where the
physics of the problem is {\it classical}.  We will be led to deal
with analogous non-perturbative problems in weakly-coupled quantum
field theory by recasting them as problems in {\it classical}
field theory.

In my example, I phrased the problem as one of high temperatures
$T$ for fixed potential $V(x)$.  However, I would have encountered the
same problem if I had held $T$ and $g$ fixed but decreased $\omega_0$:
eventually, $g^4 x^4$ would win over $\omega_0^2 x^2$.  In
gauge theory, the frequency $\omega_k$ of a field mode with
wavenumber $k$ is $\omega_k \sim k$.  So, at fixed temperature,
I may expect a problem with perturbation theory as I consider the
physics $\omega_k \sim k \to 0$ of sufficiently long wavelength
modes.  We will see that this is the case.


\subsubsection{Are RHIC QGPs weakly coupled?}

Theorists try to obtain insight by solving the problems that they can,
which are not always the problems directly relevant to experiment.
At RHIC, it is estimated that the quark-gluon plasma may have a
temperature of order $T \sim 2 T_{\rm c} \sim 350$ MeV, where
$T_{\rm c}$ is the transition temperature from a hadron gas
to a quark-gluon plasma.  So, is $\alpha_{\rm s}(\mbox{350 MeV})$
a small number?  This seems ridiculous.

\begin{figure}[th]
\centerline{\psfig{file=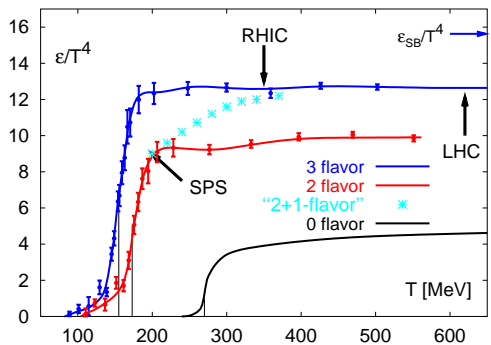,width=7cm}}
\caption{%
  Lattice data on energy density in units of $T^4$ from
  Karsch {\it et al.}\protect\cite{karsch}
  \label {fig:karsch}
}
\end{figure}

Historically, however, certain lattice QCD results lent people
optimism that weak coupling might not be such a terrible approximation
after all.  To understand this, let's focus on the energy density of
an ideal gas, where particles are not coupled at all.  First, let
me remind you of the standard Stefan-Boltzmann result for the
energy density $\epsilon$ of an ideal gas of photons, which is
\begin {equation}
   \epsilon = 2 \times \frac{\pi^2}{30} \, T^4 .
\end {equation}
Here and throughout, I work in units where $\hbar = c = \kB = 1$.
The $T^4$ comes from simple dimensional analysis.  The factor of 2 out
front represents the two possible polarizations of a photon
in a mode with a given wavenumber $\k$, and the $\pi^2/30$ is a
numerical factor that you learned to compute in a Statistical Mechanics
course.  For an ideal gas of massless pions (i.e. at temperatures
large compared to the pion mass), the result would be
\begin {equation}
   \epsilon = 3 \times \frac{\pi^2}{30} \, T^4 .
\label {eq:pion}
\end {equation}
Pions have no spin (and so no polarization) and the $3$ counts the
different types $\pi^0$, $\pi^+$, and $\pi^-$ of pions.
Similarly, for a quark-gluon plasma with massless $u$, $d$, and $s$
quarks, one gets
\begin {equation}
   \epsilon = 47.5 \times \frac{\pi^2}{30} \, T^4 .
\label {eq:sb}
\end {equation}
Here, the large factor in front counts (i) three flavors of quarks with
three colors and two spin states, (ii) the anti-quarks, and (iii) eight
colors of gluons with two polarizations.  (It's not an integer because
the fermions get a slightly different factor than $\pi^2/30$ because of
the difference between Fermi and Bose statistics.)  Now look, for
example, at the lattice data of Karsch {\it et al}.\cite{karsch} shown
in Fig.\ \ref{fig:karsch}.  It shows $\epsilon/T^4$ plotted vs.\ $T$.
At high temperature, the curve to look at is the upper one,
corresponding to three massless flavors.  The arrow labeled
$\epsilon_{\rm SB}/T^4$ in the upper-right corner shows the 3-flavor
Stefan-Boltzmann result (\ref{eq:sb}).  As you can see, it is
off by a little over 20\%.  So, from this measurement alone,
treating the system as a nearly ideal gas of quarks and gluons does
not look too bad.
The sudden jump in $\epsilon/T^4$ from low temperature to high
temperature can crudely be interpreted as a sudden jump in the number of
degrees of freedom going from a pion gas (\ref{eq:pion})
to a quark-gluon plasma (\ref{eq:sb}).

The problem, however, is that approximate ideal gas behavior is not a
sure sign of weak coupling.  There is an important
theoretical counter-example.  Instead of QCD, consider the related
theory of supersymmetric SU$(N_{\rm c})$ Yang-Mills theory with four
super-symmetries (${\cal N} = 4$).  This is a conformal theory with
vanishing $\beta$ function.  Then take the large $N_{\rm c}$ limit
($N_{\rm c} \to \infty$ with $g^2 N_{\rm c}$ fixed).  Finally, take the
infinite 't Hooft coupling limit $g^2 N_{\rm c} \to \infty$.  This
strongly-coupled four-dimensional theory has been solved using the AdS/CFT
conjecture to map it into a problem of classical gravity of a black
brane in five-dimensional anti-deSitter space.\cite{AdSCFT}
The result is
\begin {equation}
   \epsilon = 0.75 \, \epsilon_{\rm ideal} ,
\end {equation}
which is roughly the difference we saw in Fig.\ \ref{fig:karsch}.
Indeed, some people, pointing to other evidence, refer to
RHIC QGP's as strongly-coupled quark-gluon plasmas.\cite{QGL}

The moral is that one should take weak coupling with a grain of
salt for experimentally-achievable quark-gluon plasmas.  However, weak
coupling is a theoretically clean limit where one can make progress from
first principles, and one learns a lot about the physics of finite
temperature by studying it.%
\footnote{
  For attempts to apply weak coupling results to experimentally
  relevant temperatures, see Ref.\ \refcite{Tony}.
}


\section {Static Physics in Equilibrium}

\subsection {Basic equilibrium formalism}

In statistical mechanics, the basic object is the partition function $Z$,
which can be written as
\begin {equation}
  Z = \sum_{\rm states} e^{-\beta E}
    = \sum_{\rm states} \langle E | e^{-\beta H} | E \rangle
    = \operatorname{tr} e^{-\beta H} .
\end {equation}
Now note that $e^{-\beta H}$ can be written as a time evolution
operator $e^{-i Ht}$ with imaginary time $t = -i \beta$.
We know how to represent time evolution using path integrals, and
so we can represent the partition function by an imaginary-time
path integral:
\begin {equation}
   Z = \int [{\cal D}\Phi] e^{-S_{\rm E}} ,
\end {equation}
with, for example,
\begin {equation}
  S_{\rm E} = \int_0^\beta d\tau \int d^3 x
   \left[ (\partial \Phi)^2 + m^2 \Phi^2 + \lambda \Phi^4 \right]
\end {equation}
and with {\it periodic}\/ boundary conditions
\begin {equation}
  \Phi(0,x) = \Phi(\beta,x)
\end {equation}
in imaginary time for bosonic fields.
These periodic boundary conditions implement the trace in
$Z = \operatorname{tr} e^{-\beta H}$.  For fermionic fields, it turns out that
(to implement Pauli statistics) one must impose
anti-periodic boundary conditions.%
\footnote{
  For textbook treatments in the context of relativistic field theory,
  see Ref.\ \refcite{textbooks}.
}

Feynman rules are exactly the same as in zero-temperature field theory
except that the imaginary time $\tau$ is now periodic with period
$\beta$.  To go from $\tau$ to frequency space, this means we should
do a Fourier series rather than a Fourier transform.
The only difference with zero-temperature Feynman rules will then
be that loop
frequency integrals are replaced by loop frequency sums:
\begin {equation}
  \int \frac{d^4P}{(2\pi)^4} \rightarrow
  T \sum_\nu \int \frac{d^3p}{(2\pi)^3}
\end {equation}
with the sum over discrete imaginary-time frequencies
(known as Matsubara frequencies)
\begin {align}
  \nu_n &= 2\pi n/\beta = 2\pi n T & \mbox{(bosons)} \\
  \nu_n &= 2\pi(n+\tfrac12)/\beta = 2\pi(n+\tfrac12)T & \mbox{(fermions)}
\end {align}
to implement the periodic or anti-periodic boundary conditions.


\subsubsection {An example}

\begin{figure}[th]
\centerline{\psfig{file=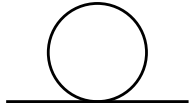,width=2.5cm}}
\caption{%
  A 1-loop contribution to the effective mass.
  \label {fig:loop}
}
\end{figure}

\begin{figure}[th]
\centerline{\psfig{file=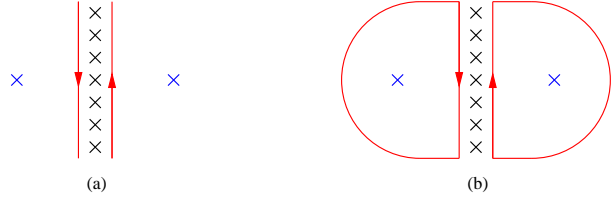,width=8cm}}
\caption{%
  Integration contour for doing frequency sum..
  \label {fig:contour}
}
\end{figure}

As an example, consider the self-energy diagram of Fig.\
\ref{fig:loop}, which contributes to the effective mass of
a particle.  This loop gives
\begin {equation}
   \Delta m_{\rm eff}^2
   = \frac{\lambda}{2} \, T \sum_{\nu} \int \frac{d^3 p}{(2\pi)^3} \>
       \frac{1}{\nu^2+p^2+m^2} \,.
\label {eq:sum}
\end {equation}
We can evaluate the sum by noticing that the Bose distribution
function
\begin {equation}
   n_{\rm B}(\omega) = \frac{1}{e^{\beta\omega} - 1 }
\end {equation}
has simple poles with residue $T$ at $\omega = i 2\pi n T = i \nu_n$.
So use the Residue Theorem to rewrite the sum (\ref{eq:sum}) as
\begin {equation}
   \Delta m_{\rm eff}^2
   = \frac{\lambda}{2} \int_C \frac{d\omega}{2\pi i}
       \int \frac{d^3 p}{(2\pi)^3} \>
       \frac{n_{\rm B}(\omega)}{-\omega^2+\omega_p^2} \,,
\end {equation}
where the contour $C$ is shown in Fig.\ \ref{fig:contour}a and
$\omega_p^2 \equiv p^2 + m^2$.  Now close each line of the contour at
infinity to enclose the poles of the integrand at
$\omega = \pm \omega_p$ as in Fig.\ \ref{fig:contour}b.
Again using the Residue Theorem,%
\footnote{
  It should be easy to see how the $\omega=+\omega_p$ residue works out.
  The $\omega=-\omega_p$ residue generates a similar result
  using the identity
  $n_{\rm B}(-\omega_p) = -[1 + n_{\rm B}(\omega_p)]$.
}
\begin {equation}
   \Delta m_{\rm eff}^2
   = 
       \int \frac{d^3 p}{(2\pi)^3 2\omega_p} \>
       \lambda \, n_{\rm B}(\omega_p)
   \quad + \quad
   \mbox{(a $T{=}0$ contribution)} .
\label {eq:loop}
\end {equation}

There is a simple physical way to understand the result in terms of
the propagation of a particle through the thermal medium.
Eq.\ (\ref{eq:loop}) can be rewritten in the form
\begin {equation}
  \raisebox{-0.3cm}{\psfig{file=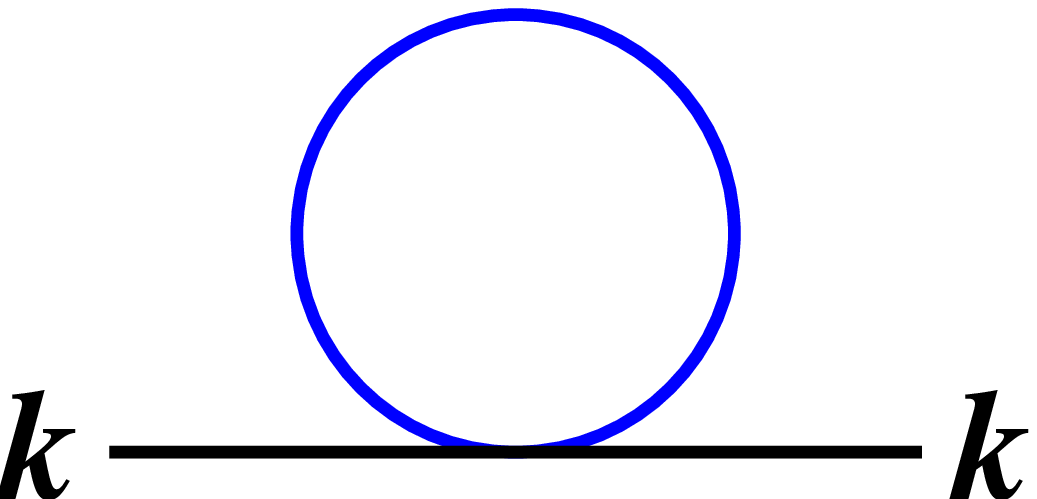,width=1.5cm}}
  =
  \raisebox{-0.3cm}{\psfig{file=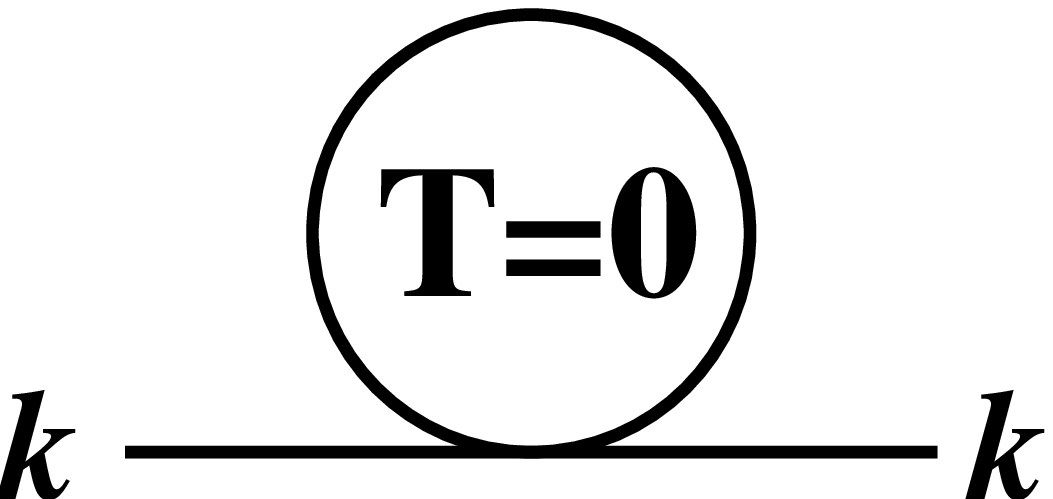,width=1.5cm}}
  - \int \frac{d^3 p}{(2\pi)^3 2 \omega_p} \, n_{\rm B}(\omega_p)
  \raisebox{-0.3cm}{\psfig{file=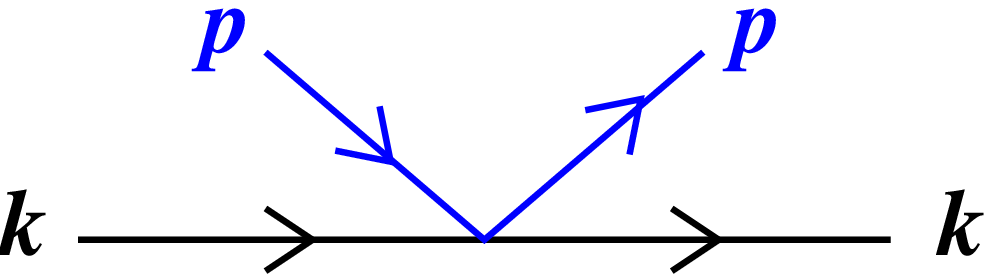,width=2.5cm}}
  .
\end {equation}
The last term is the
amplitude for forward scattering of the particle of interest
(of momentum $k$) off of a particle in the
plasma (of momentum $p$), integrated over relativistic phase
space, weighted by the probability density $n_{\rm B}(\omega_p)$
for encountering such a particle in the plasma.
In this case, the amplitude is just a factor of $\lambda$.

The moral is that the thermal contribution to the effective mass
comes from forward scattering off of the medium.
This is similar to
the index of refraction for light in a medium, which is also
caused by forward scattering and which slows light below $c$,
like a mass does.

For future reference, note that at high temperatures ($T \gg m$), the
integral in (\ref{eq:loop}) is dominated by $p \sim T$, which is the
momentum of typical particles in the thermal gas.


\subsection{How big are loops?}

Because the integral in (\ref{eq:loop}) is dominated by
momenta $p \sim T$, the result, by
dimensional analysis, is
\begin {equation}
   \Delta m_{\rm eff}^2 = \# \lambda T^2
\end {equation}
in the high-temperature limit.
The numerical coefficient is something you can easily
compute by doing the integral.

For high temperatures, this loop correction to the effective mass
can have a profound effect.
Consider the example of high-temperature symmetry restoration.
Suppose we start with a potential
\begin {equation}
  V(\phi) = -\mu^2 \phi^2 + \lambda \phi^4
\end {equation}
at zero temperature.  The minima of this potential are at
non-zero $\phi$, resulting in spontaneous symmetry breaking.
However, the thermal correction to the mass will replace
the $-\mu^2$ by an effective $-\mu^2 + \# \lambda T^2$:
\begin {equation}
  V(\phi) \to (-\mu^2 + \# \lambda T^2) \phi^2 + \lambda \phi^4 .
\end {equation}
For
\begin {equation}
  T > T_{\rm c} = \frac{\mu}{(\# \lambda)^{1/2}} \,,
\end {equation}
the coefficient of $\phi^2$ is now positive, the minimum is at
$\phi=0$, and so thermal effects have restored the symmetry at
high temperature.

The moral is that, at large enough temperature, loop effects can
be large!

If loop effects are large, how can we use perturbation theory?
The one-loop diagram just analyzed turns out to be the worst
offender.%
\footnote{
  For a discussion of the size of loop effects, see the original
  papers on the subject.\protect\cite{loops}
}
So let's resum it into propagators, replacing
our perturbative propagator $(P^2+m_0^2)^{-1}$ with
$(P^2 + m_{\rm eff}^2)^{-1}$, where
\begin {equation}
  m_{\rm eff}^2 = m_0^2 + \# \lambda T^2 .
\end {equation}

\begin{figure}[th]
\centerline{\psfig{file=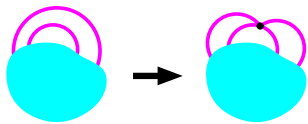,width=4cm}}
\caption{%
  Generically adding another loop to a diagram.
  \label {fig:addloop}
}
\end{figure}

That takes care of the loops like Fig.\ \ref{fig:loop}, but then
what does it cost to add other sorts of loops to a diagram
as in Fig.\ \ref{fig:addloop},
when $T$ is large?
The cost is an explicit factor of coupling, a new loop integration,
and two new propagators:
\begin {subequations}
\label {eq:addloop}
\begin {equation}
  \lambda \times T \sum_\nu \int d^3p \times \mbox{(two propagators)} .
\end {equation}
There's an explicit factor of $\lambda T$ above, and then the rest
has dimensions on inverse momentum:
\begin {equation}
   \mbox{cost} \sim \frac{\lambda T}{\mbox{momentum}} \,.
\end {equation}
\end {subequations}
The biggest cost would therefore seem to be due to the most infrared
momentum.  In the case at hand, the infrared is cut off by the
effective mass $m_{\rm eff}$, so%
\footnote{
   The logic here assumes that the added three-dimensional loop
   integral is dominated by the infrared.  That's generically true,
   but not for insertions of Fig.\ \ref{fig:loop}, which gives a
   divergent sub-diagram that leads to an extra factor of $T$ in
   its thermal contribution.
   But we've resummed all of those diagrams.  There are also
   logarithmically-divergent sub-diagrams, which can give additional
   factors of $\ln(T/m)$.
}
\begin {equation}
   \mbox{cost} \sim \frac{\lambda T}{m_{\rm eff}}
   \sim \frac{\lambda T}{(\lambda T^2)^{1/2}}
   \sim \sqrt\lambda .
\end {equation}

So, for scalar theory, the small coupling expansion at high $T$
is an expansion in $\sqrt\lambda$.  As an example, the
free energy is known to high order and has the form\cite{Fscalar}
\begin {equation}
  P = T^4 [1 + \# \lambda + \# \lambda^{3/2} + \# \lambda^2
             + \lambda^{5/2} (\# \ln\lambda + \#)
             + \lambda^3 (\# \ln\lambda + \#)
             + \cdots] .
\end{equation}
The $\ln\lambda$ comes from a logarithm $\ln(T/m_{\rm eff})$
that arises in certain diagrams.

For scalar theory, it is enough to resum the effects of Fig.\
\ref{fig:loop} into propagators to make a workable
perturbation theory and, in principal, find the expansion
to as high an order as one has the will to go.  We'll see
later that gauge theory is different.


\subsection{Another way to understand $\lambda T/m$}

Here's another, more physical way to understand why the cost of
interactions is $\lambda T/m_{\rm eff}$ and not simply $\lambda$.
Consider the limit of the Bose distribution function for small
particle energies $E \ll T$:
\begin {equation}
  n_{\rm B}(E) = \frac{1}{e^{\beta E}-1} \to
  \frac{1}{\beta E} = \frac{T}{E} .
\end {equation}
This is largest when $E \sim m_{\rm eff}$, giving
\begin {equation}
  n_{\rm B} \sim \frac{T}{m_{\rm eff}} \,.
\label {eq:nBlimit}
\end {equation}
Now think about adding a loop, such as
\begin {equation}
  \raisebox{-0.5cm}{\psfig{file=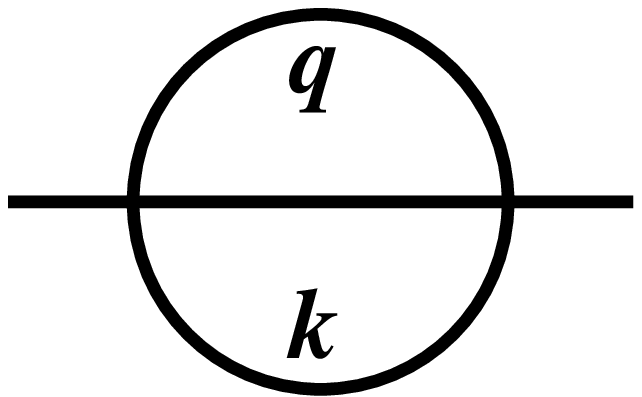,width=2cm}}
  \longrightarrow
  \raisebox{-0.5cm}{\psfig{file=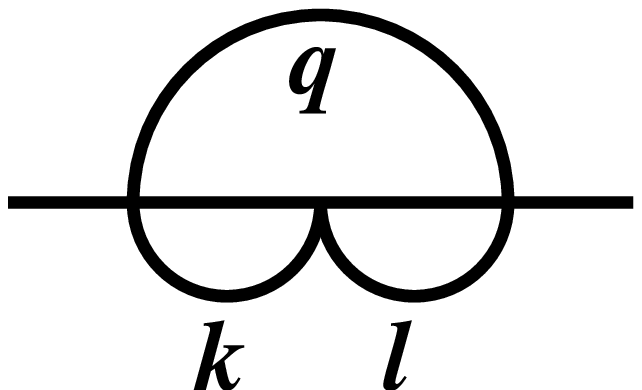,width=2cm}}
  .
\end {equation}
In terms of forward scattering off of the medium, adding an
extra interaction corresponds, for instance, to
\begin {equation}
  \raisebox{-0.5cm}{\psfig{file=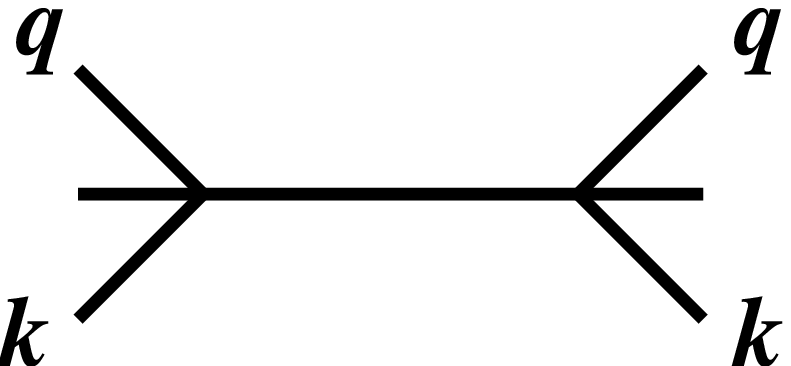,width=2.5cm}}
  \longrightarrow
  \raisebox{-0.5cm}{\psfig{file=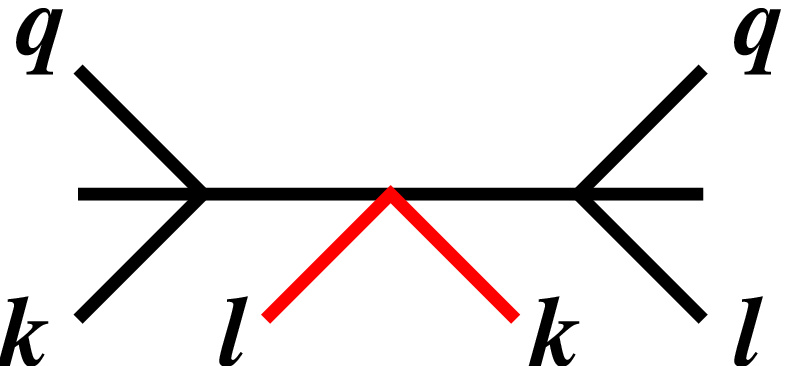,width=2.5cm}}
  + \cdots
  .
\end {equation}
What is the cost?  There is an explicit $\lambda$ for the
extra interaction, but there is also a factor of density
$n_{\rm B} \sim T/m_{\rm eff}$
for the probability of an additional particle from
the medium being involved.  So the total cost is
\begin {equation}
   \lambda n_{\rm B} \sim
   \frac{\lambda T}{m_{\rm eff}} \sim \sqrt{\lambda} .
\end {equation}
In this language, we see that the $T/m_{\rm eff}$ factor is
an enhancement due to the large density
(\ref{eq:nBlimit}) of low-momentum particles
in a Bose gas.  (There is no similar enhancement 
for fermions.)


\subsection{Gauge bosons ``masses''}
\label {sec:gmass}

\begin{figure}[th]
\centerline{\psfig{file=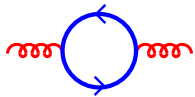,width=2.5cm}}
\caption{%
  A contribution to the 1-loop self-energy for gluons.
  \label {fig:gloop}
}
\end{figure}

Now I'll turn to gauge theory.  Suppose one evaluates one-loop
self-energies like Fig.\ \ref{fig:gloop}, plus similar diagrams
with gauge boson loops.  Let me consider the propagation of
relatively soft gauge bosons, with $\omega$ and $k$ small compared
to $T$.  It turns out the self-energy then has the functional
form\cite{textbooks,pi}
\begin {equation}
  \Pi_{\mu\nu}(\omega,k) = g^2 T^2 \, f_{\mu\nu}(\omega/k) .
\label {eq:piform}
\end {equation}
Overall, its size is determined by $g^2 T^2$, analogous to the
$\lambda T^2$ we discussed in the scalar case.  Unlike the
scalar theory,
the diagram of Fig.\ \ref{fig:gloop} is sensitive to the
external momentum, and the result turns out to
depend on the ratio $\omega/k$ (and the direction of $\k$).
I'm not going to give details.  Instead, I will just tell you
some interesting limiting cases and explain how the results
make physical sense.

\subsubsection {Debye screening}
\label {sec:debye}

Let's consider the case $\omega=0$ (static response).
It turns out that
\begin {equation}
  \Pi^\mu_\nu
  = \mD^2 \begin{pmatrix} 1&&&\\&0&&\\&&0&\\&&&0 \end{pmatrix} ,
\label {eq:Pistatic}
\end {equation}
where
\begin {equation}
  \mD^2 = (1+\tfrac16 N_{\rm f}) g^2 T^2
\end {equation}
and $N_{\rm f}$ is the number of quark flavors.  If we sum up insertions
of this
self-energy into the gauge propagator, the propagator
for the electric potential $A^0$ at $\omega=0$ is $(p^2+\mD^2)^{-1}$.
If we 3-dimensionally Fourier transform this static propagator, it
should be proportional to the electric potential between two static
test charges.  Because of the effective mass $\mD$, this Fourier
transform gives a Yukawa potential $e^{-\mD r}/r$ instead of the usual
Coulomb $1/r$.  This effect is well known in plasmas and is called
Debye screening.  It's something you can find discussed in detail
in classical terms in Jackson.\cite{Jackson}
In our case, the
corresponding screening length $\xiD$ is
\begin {equation}
  \xiD = \frac{1}{\mD} \sim \frac{1}{gT} .
\end {equation}

Another way to understand color deconfinement at high temperature
(i.e.\ the quark-gluon plasma as a nearly free gas of quarks and gluons)
is in terms of Debye screening.  Start by placing
a negative static test charge in a QED plasma.  Debye screening is
caused by positive charges flying by and being bent slightly towards the
test charge, while negative charges are repelled slightly away from it,
so that on average there is a cloud of positive screening charge around
the negative test charge.  The radius of this cloud is the Debye
screening length $\xiD$, which, in both relativistic and
non-relativistic situations, is order
\begin {equation}
  \xiD \sim \sqrt{\frac{T}{g^2 n}} ,
\label {eq:xiD}
\end {equation}
where $n$ is the density of charged particles in the plasma.  Outside
of this radius, the average electric field falls exponentially with
distance from the test charge.  There is a critically important
difference to notice between ultra-relativistic and non-relativistic
plasmas.  In the non-relativistic case, the density $n$ is
fixed, and so the Debye screening length (\ref{eq:xiD})
{\it increases}\/ with temperature.
(It's harder to bend particles that are more energetic.)
In the ultra-relativistic
case, $n \sim T^3$ increases with temperature, due to pair creation.
And this increase is rapid enough that the screening length
(\ref{eq:xiD}) {\it decreases}\/ with increasing temperature.
(There are more particles to participate in screening.)
As a result, electric forces only operate over small distances
at high temperature.  Now consider a non-Abelian plasma.  At
zero temperature, the potential between two static test charges
looks Coulomb ($\alpha/r$) at very short distances [with a running
$\alpha = \alpha(r^{-1})$], but is confining (linear in $r$) at
large distances.  But the fact that Debye screening only allows
electric forces to operate over small distances in the limit of high,
ultra-relativistic temperatures means that, if the temperature is
large enough, the long-distance confining behavior will be
screened away.


\subsection{Magnetic fields}

We saw the Debye effect manifest in (\ref{eq:Pistatic})
as a mass $\mD$ for the electric potential $A^0$.
Note that there is no similar mass for the magnetic potential
$\A$ in (\ref{eq:Pistatic}).  This reflects the fact that, unlike
electric fields, plasmas do not screen static magnetic fields.
It's the reason why the Sun and the galaxy can support magnetic fields.
It's the reason that magnetic forces can act over long distances and
create complicated structure like solar filaments lifting off of
the surface of the sun, and why there is a whole field of
classical electromagnetism called magneto-hydrodynamics (MHD).

In QED plasma, there is no screening of static magnetic fields.
What happens in QCD?  In the absence of screening, one might wonder
if ordinary confinement once again rears its head
at sufficiently large distances.  Maybe a QCD plasma has electric
deconfinement (because of the Debye effect) but confinement of
magnetic fields (because of the absence of a similar effect).
This turns out to be the case, but the physics and scale
of ``magnetic confinement'' is quite different than the
more familiar confinement of zero temperature QCD at 1 fermi.

Confinement is a non-perturbative phenomenon.
At what range do QCD magnetic forces become non-perturbative at
high temperature?
One way to investigate is
to think about attempting perturbation
theory and determine the generic cost of adding each additional
loop, just as previously considered for scalar theory
in Fig.\ \ref{fig:addloop}.
The answer is just like (\ref{eq:addloop}), but with $\lambda$
replaced by $g^2$:
\begin {equation}
  \mbox{cost}
  \sim g^2 \times T \sum_\nu \int d^3p \times \mbox{(two propagators)}
  \sim \frac{g^2 T}{\mbox{momentum}} \,.
\label {eq:QCDaddloop}
\end {equation}
But now there is a huge difference between gauge theory and scalar
theory.  In scalar theory, there was an effective mass to cut off infrared
momentum.  In gauge theory, in the static case, there is a mass
for the electric potential $A^0$ but not for the magnetic
potential $\A$.  Let's restrict attention to the contribution from
$\nu=0$ in the frequency sum in (\ref{eq:QCDaddloop}), for which
the static case is relevant.  Because there is no mass to cut off
magnetic gluons in the infrared, (\ref{eq:QCDaddloop}) is infrared
divergent.  The cost $g^2T/p$ of adding a loop becomes
non-perturbative, that is $\gtrsim O(1)$, when the momentum scale
in (\ref{eq:QCDaddloop}) is $p \lesssim g^2 T$.  The scale of
non-perturbative magnetic physics is therefore
\begin {equation}
  p_{\rm mag} \sim g^2 T .
\end {equation}
The origin of this non-perturbative physics is just like that for
the simple quantum-mechanical example I gave back in
Sec.\ \ref{sec:QMexample}.

I have referred to the non-perturbative magnetic physics as
magnetic ``confinement.''  Why do I call it confinement?
To explain, it will be helpful to first discuss a little more
formalism.


\subsubsection {Imaginary-time dimensional reduction}
\label{sec:dimred}

For simplicity, let's return for the moment to scalar theory.  Consider
the imaginary time action
\begin {equation}
  S_{\rm E} = \int_0^\beta d\tau \int d^3 x
   \left[ (\partial \Phi)^2 + \lambda \Phi^4 \right]
\label {eq:SE}
\end {equation}
at finite temperature, where I've assumed the temperature is high
enough that I can ignore the zero-temperature mass term.
This action is integrated over a slab
of Euclidean space that is spatially infinite but
extends only for $\beta$ in the time direction, with periodic boundary
conditions.  In the limit of higher and higher temperature, this slab is
very thin and looks more and more three-dimensional.  Imagine looking at
the slab with blurred eyes, so that you can only resolve physics
on distance scales large compared to $\beta$.  You would
then be unable to tell the difference between a slab of thickness
$\beta$ and one of thickness zero.  That is, the effective theory
which described long distance physics (distances $\gg \beta$) is
a {\it three}-dimensional Euclidean field theory.

A naive version of this is to formulate the three-dimensional
version by considering only static modes above, so that
$\Phi(\tau,x) \to \Phi(x)$.  That is, the dynamics of the tiny
imaginary time direction decouples if we look at long-distance physics.%
\footnote{
  just like excited Kaluza-Klein modes decouple in a theory with tiny,
  compact extra dimensions.
}
The $d\tau$ integral in (\ref{eq:SE}) then becomes trivial,
giving a factor of $\beta$, so that
\begin {equation}
  S_{\rm E} \to S_3
  = \beta \int d^3 x
   \left[ (\nabla \Phi)^2 + \lambda \Phi^4 \right]
  = \int d^3 x
   \left[ (\nabla \phi)^2 + \lambda T \phi^4 \right] ,
\end {equation}
where I've defined $\phi \equiv \beta^{1/2}\Phi$ just
to give the kinetic term of the three-dimensional theory a conventional
normalization.

In more detail, saying that the imaginary time dynamics decouples is
not quite the same as saying it has no effect: it can give finite
renormalizations to the interactions of static modes
when one integrates out the
non-static ($\nu_n \not = 0$) modes of the field.
In particular, consider the contribution to
the one-loop self-energy diagram of
Fig. \ref{fig:loop} when the external lines are static
$(\nu=0)$ but the internal line is summed over the non-static
$(\nu_n\not = 0)$ modes which are decoupling.  This contribution
turns out to give the $\#\lambda T^2$ thermal contribution to
$m_{\rm eff}^2$ that we have discussed earlier.  So, the decoupling
imaginary time dynamics contributes to give an effective mass
in the long-distance three-dimensional effective theory of
\begin {equation}
  m_3^2 = m_0^2 + \# \lambda T^2 .
\end {equation}
This is the most important effect.  There are also sub-leading effects
on the coupling $\lambda_3 \simeq \lambda T$ of the three-dimensional
theory and the generation of sub-leading higher-dimensional
interactions, such as $\nu_3 \phi^6$ with $\nu_3 \sim \lambda^3$.
Perturbative calculations can be done to determine all the
coefficients of interest in the effective theory, and such
imaginary-time ``matching calculations'' are a mature, well-understood
technology.

What happens if we ask the same question about gauge theory?
If we do naive dimensional reduction,
the 4-dimensional action at high temperature
\begin {equation}
  S_{\rm E}
  = \int_0^{\hbar\beta} d\tau \int d^3 x \> F^2
\end {equation}
reduces to
\begin {equation}
  S_{\rm E}
  = \beta \int d^3x \>  F^2 ,
\end {equation}
and the $\beta$ can again be absorbed by a field redefinition, making
the three-dimensional coupling $g_3^2 = g^2 T$.
Ignore $A^0$ for a moment.  The
moral is then that the effective long-distance (i.e. low-momentum)
theory for high temperature four-dimensional gauge theory is
(zero temperature) {\it three}-dimensional gauge theory,
with the scale set by $g^2 T$.  The
question about whether the long-distance physics is confining is then a
question about whether three-dimensional non-abelian Yang-Mills is
confining.  The answer (from lattice simulations) turns out to be that
it is.%
\footnote{
  What happens to the $A^0$?  It picks up the Debye mass from
  integrating out the effects of the non-zero frequency modes.
  In three-dimensional language, it becomes a massive, adjoint color
  scalar field and, because of its mass,
  decouples at distances $\gg 1/\mD \sim 1/gT$.
}

It's interesting to put explicit $\hbar$'s back into this discussion.
The dimensional reduction is then
\begin {equation}
  \frac{S_{\rm E}}{\hbar}
  = \hbar^{-1}\int_0^{\hbar\beta} d\tau \int d^3 x \> F^2
  \to \hbar^{-1} \times \hbar\beta \int d^3x \>  F^2
  = \beta \int d^3x \>  F^2 ,
\label {eq:SE2}
\end {equation}
where the $\hbar$ has to appear in the upper limit of the $d\tau$
integration by dimensional analysis if we define $\beta \equiv 1/k_{\rm B}T$.
The absence of $\hbar$ on the far right side of (\ref{eq:SE2}) shows
that the effective long-distance theory is a
{\it classical}\/ field theory, consistent with the discussion
in Sec.\ \ref{sec:QMexample} that non-perturbative physics can
occur at high temperatures in weakly coupled theories but that it
will be classical in nature.


\subsection{Summary of Scales}
\label {sec:scales1}

\begin{figure}[th]
\centerline{\psfig{file=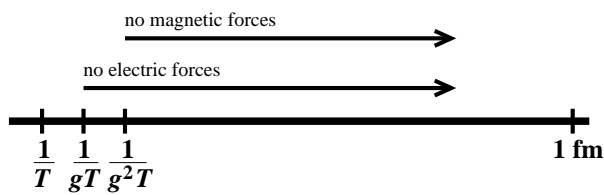,width=8cm}}
\caption{%
  A cartoon of the distance scales relevant to time-independent
  questions of equilibrium, high-temperature gauge theory.
  \label {fig:scales1}
}
\end{figure}

We've now seen the following hierarchy of momentum scales:
\begin {itemlist}
\item
$\phantom{g^2} T$: typical particle momenta;
\item
$\phantom{{}^2}gT$: Debye mass, causing charge deconfinement;
\item
$g^2 T$: magnetic confinement.
\end {itemlist}
The corresponding distance scales are shown in
Fig.\ \ref{fig:scales1} .

\section {Dynamics Near Equilibrium: Plasma Waves}

For $\omega$ and $k$ both small compared to $T$, I previously
asserted that the gauge boson self-energy has the form
$\Pi_{\mu\nu}(\omega,k) = g^2 T^2 \, f_{\mu\nu}(\omega/k)$.
So far, I have only discussed the static limit $\omega=0$
(applying to $\omega \ll k \ll T$),
where the self-energy was given by (\ref{eq:Pistatic}).
Let's now consider the opposite limit $k=0$
(applying to $k \ll \omega \ll T$).
This is not a natural limit in the imaginary-time formalism,
where $\omega = i \nu_n = 2\pi n T$.  Imaginary frequencies cannot
be simultaneously non-zero and small compared to $T$.
But one can find the desired result by appropriate analytic continuation.
It turns out to give\cite{textbooks,pi}
\begin {equation}
  \Pi^\mu_\nu
  = \mD^2
  \begin{pmatrix} 0&&&\\&\tfrac13&&\\&&\tfrac13&\\&&&\tfrac13 \end{pmatrix} .
\end {equation}
This generates a mass gap for low-momentum (color) electromagnetic
waves propagating in the plasma:
\begin {equation}
  m_{\rm pl} = \frac{\mD}{\sqrt3} \,.
\end {equation}
This is known as the plasma frequency.  (See Jackson!\cite{Jackson})
Or, regarding plasma wave quanta, it is known as the
soft plasmon mass.

\begin{figure}[th]
\centerline{\psfig{file=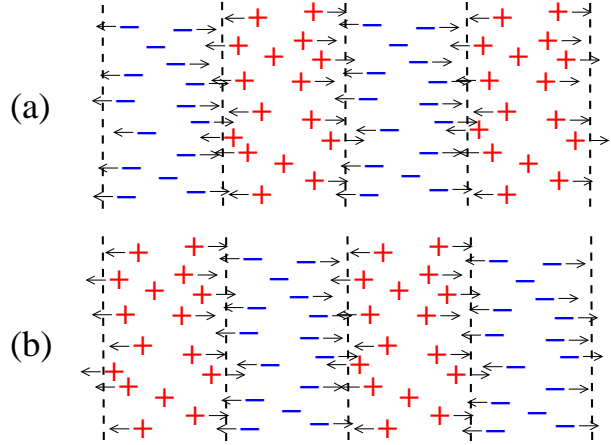,width=8cm}}
\caption{%
  \label {fig:longitudinal}
  A picture of a longitudinally polarized plasma wave as
  oscillations between (a) and (b).
}
\end{figure}

A massive gauge field has a longitudinal polarization, in addition to
the two transverse polarizations familiar from propagating
electromagnetic waves in vacuum.  The physical picture of a
longitudinal plasma wave is that it is a charge wave together with
the associated electric fields.
For simplicity, consider QED rather than QCD.
Suppose that an equilibrium plasma
is disturbed to have a small excess of charge in certain regions,
as depicted in Fig.\ \ref{fig:longitudinal}a (which shows only the
excess charge).  Electric forces will cause the charges to accelerate
in the directions shown by the arrows in that figure.
The charges will then move to cancel each other, except they will
overshoot because of inertia, leading to the situation shown
in Fig.\ \ref{fig:longitudinal}b, with electric forces again shown
by the arrows.  The wave will oscillate between these two states
(eventually dissipating because of random collisions---a subleading
effect).%
\footnote{
  Note that ultra-relativistic particles move at nearly the speed
  of light and so do not simply stop and turn around.  Moreover,
  my discussion of the self-energy $\Pi$ has been perturbative, and
  small, perturbative fields can only have small effects on the
  otherwise straight-line trajectories of the particles.  These
  small deviations, however, can slightly increase the density
  of one charge type in one place and decrease it in another.
  Fig.\ \ref{fig:longitudinal} should be considered a
  particle-averaged picture of local charge fluctuations, rather
  than a description of the motion of each individual particle.
}

In general, the dispersion relation for propagating gluons in the
plasma has a complicated form but two simple limiting cases:
\begin {align}
  k \ll m_{\rm pl}:& \qquad \omega^2 \simeq k^2 + m_{\rm pl}^2
  & \mbox{(transverse, longitudinal)}
\\
  k \gg m_{\rm pl}:& \qquad \omega^2 \simeq k^2 + \tfrac12 m_{\rm pl}^2
  & \mbox{(transverse)}
\end {align}
The first case is the one just discussed, which we'll call the soft
gluon case.  The second case says
that ``hard'' gluons have a mass of
$m_\infty \equiv m_{\rm pl}/\sqrt2 = \mD/\sqrt{6} \sim gT$.

Fermions gets similar masses of order $gT$ from medium
effects on their propagation.  The moral is that, at very
high temperature, all propagating (quasi-)particles get masses
of order $gT$.

This might seem contradictory.  If all propagating particles have
masses of order $gT$, then what were those ``massless'' low-momentum
($k \sim g^2 T$) gluons that I said caused magnetic confinement?
They came from the spatial polarizations of the
$\omega=0$ ($\omega \ll k \ll T$) self-energy
(\ref{eq:Pistatic}).
They are not independently propagating excitations because they have
{\it space}-like ($\omega <k$) momenta.
What they represent are not propagating particles themselves (which have
time-like momenta) but the magnetic fields created by those
moving particles.


\section {Dynamics Near Equilibrium: Relaxation and Viscosity}

Previously, I have emphasized the use of imaginary time path
integrals in understanding various static aspects of thermal
equilibrium.  If one is studying static properties (the
equation of state, equal-time correlation functions, and so forth),
it doesn't matter whether one works in imaginary time or real time.
And some things, like magnetic confinement, are easiest to see
and understand in the imaginary time formalism.
For static quantities, the imaginary time formalism, dimensional
reduction, and perturbative matching calculations have provided
a systematic, mature technology for studying problems order by
order in small coupling.

\begin{figure}[th]
\centerline{\psfig{file=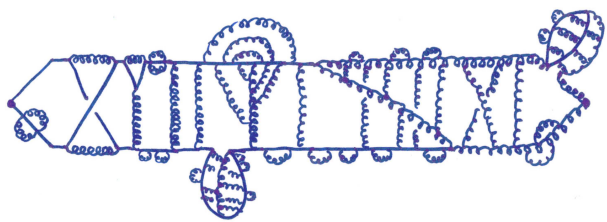,width=8cm}}
\caption{%
  \label {fig:mess}
  One of an infinite class of graphs contributing to the
  QCD shear viscosity at leading-order (except for an
  obvious error).
}
\end{figure}

In contrast, there are many time-dependent aspects of plasmas (such as
viscosity, electrical conductivity, and other quantities characterizing
relaxation) where it is a struggle to get even leading-order results.
Organizing small-coupling calculations in some more useful way than the
traditional loop expansion turns out to be really important, as
exemplified by Fig.\ \ref{fig:mess}.  This figure depicts just
one of an infinite set of diagrams that contributes to the
{\it leading}-order result for the shear viscosity.
(Actually, there is an obvious error in the diagram which I have
not corrected.  Can you spot it?\footnote{
  Answer: The long diagonal gluon line was meant to be a quark line.  There is
  no such thing as a quark-gluon-gluon vertex.
}%
)

Indeed, there are basic aspects of the dynamics of plasmas which
are incredibly difficult to see in imaginary time.


\subsection {A failure of imaginary-time intuition}

For simplicity, let's return to scalar theory.
In Sec.\ \ref{sec:dimred}, we learned that the long-distance effective
theory is a three-dimensional theory
\begin {equation}
   S_3 = \int d^3 x \left[ (\nabla \phi)^2 + m_3^2 \phi^2
           + \lambda_3 \phi^4 + \cdots \right]
\end {equation}
where $\phi \simeq \beta^{1/2} \Phi$,
$m_3^2 \simeq m_0^2 + \# \lambda T^2$,
and $\lambda_3 \simeq \lambda T$.
There is a non-zero effective mass $m_3$ in this theory.
Technically, this means that is we evaluate equal-time correlators,
we'll find that they fall exponentially at large distances:
\begin {equation}
  \langle \phi(0,\x) \, \phi(0,\y) \rangle \sim e^{-m_3|\x-\y|} .
\end {equation}

\begin{figure}[th]
\centerline{\psfig{file=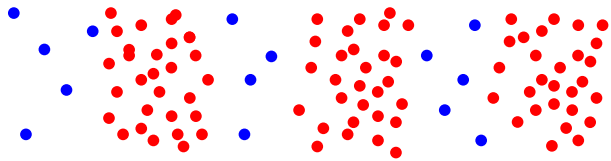,width=8cm}}
\caption{%
  \label {fig:sound}
  A cartoon of a sound wave in an ultra-relativistic gas.
}
\end{figure}

The presence of a mass in the above low-momentum effective theory
might make you think that there is a minimal
energy $m_3$ for creating excitations of the system.
This would be wrong.  At finite temperature, there are always
long wavelength, low frequency oscillations in the form
of sound waves, such as depicted (for an
ultra-relativistic theory) in Fig.\ \ref{fig:sound}.
A sound wave is a pressure wave.  Higher pressure means higher energy
and so, in the ultra-relativistic limit, higher particle density
(because of pair creation).  So, in the cartoon of Fig.\
\ref{fig:sound}, I've indicated regions of higher pressure by
drawing more particles.  If the wavelength is long, the period of
oscillation has to be long, because energy is conserved: the
only way that high and low energy regions can swap places is for
energy to be transferred over distances of order a wavelength,
and the rate of such transfer is limited at the very least
by the speed of light.
In general, a sound wave has
\begin {equation}
   \omega \simeq v_{\rm s} k \to 0
   \qquad \mbox{as} \qquad
   k \to 0 ,
\end {equation}
where $v_{\rm s}$ is the speed of sound.

So what was wrong with thinking that the mass $m_3$ in the
three-dimensional effective theory implied an energy gap (and so
minimum frequency) for excitations?  This is a problem of thinking too much
in imaginary time.  The three-dimensional high-temperature effective
theory is an effective theory of low-momentum {\it static}\/ physics
($\omega=0$ and $k \ll T$).  But sound waves have $\omega \sim k \ll T$.


\subsection {A sequence of real-time effective theories}

Sound waves are an example of a hydrodynamic phenomenon.
In general, if I am interested in the real-time physics
of a finite temperature system at long enough distance and time
scales, I can describe the system using hydrodynamics.
Over long time scales, the only excitations that will not
have decayed away are long wavelength variations in locally
conserved quantities, such as energy and momentum density.
(Because of conservation, these quantities can only change
as fast as they can move from one place to another, as I just
described for sound.)  The effective theory of such quantities
is hydrodynamics.%
\footnote{
  A nice place to very quickly pick up the basics of non-relativistic
  hydrodynamics is chapters 41--42 of the Feynman lectures,
  volume 2.\cite{Feyn2}
  For relativistic hydro, try Weinberg.\cite{Weinberg}
}

What if we want to compute, from first principles, parameters of
the hydrodynamic description, such as viscosity (to be reviewed
shortly)?  It turns out it's somewhat overwhelmingly complicated
to jump in a single step from the microscopic description in
terms of quantum field theory to the macroscopic description in
terms of hydrodynamics.  Diagrams like Fig.\ \ref{fig:mess} are
involved.  What's needed is to organize the calculation by
stepping first through another effective theory at intermediate
(and longer) scales.  That modern, state-of-the-art
effective theory was just invented... in 1872, by some fellow named
Boltzmann.  It is kinetic theory.


\subsubsection{The form of the Boltzmann equation}

In kinetic theory, one describes the state of the system in terms of
a classical phase space density of particles, $f(\x,\p,t)$.
The basic equation of kinetic theory is
simply the statement that (in the absence of external forces)
changes in momentum $\p$ only occur by loss or gain
through scattering:
\begin {equation}
  \frac{d f_\p}{dt} = - (\mbox{loss})_\p + (\mbox{gain})_\p .
\end {equation}
For example, schematically,
\begin {equation}
   \frac{\partial f}{\partial t}
   + \frac{\partial\x}{\partial t} \cdot \frac{\partial f}{\partial\x}
   = - \biggl| 
   \raisebox{-0.2cm}{\psfig{file=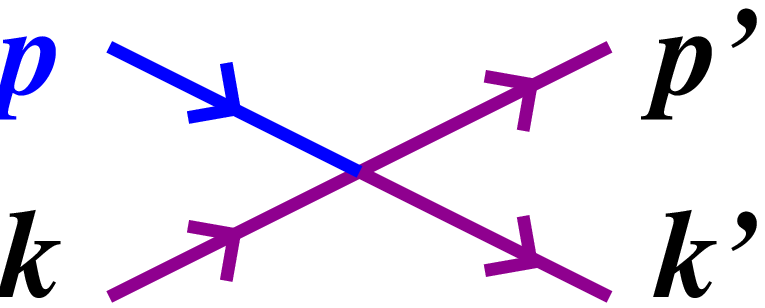,width=1.5cm}}
   \biggr|^2 + \biggl|
   \raisebox{-0.2cm}{\psfig{file=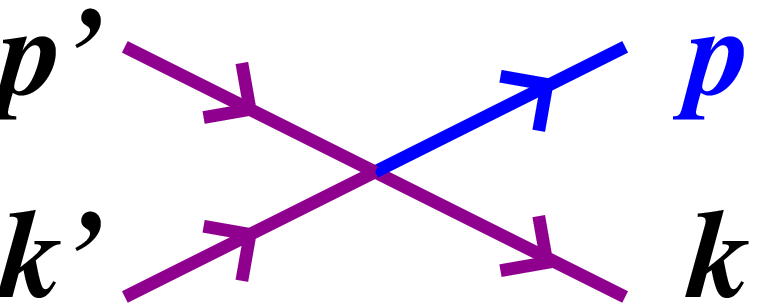,width=1.5cm}}
   \biggr|^2 .
\label {eq:boltz0}
\end {equation}
Here, I've just used the chain rule on the left-hand side.
I've represented the loss and gain terms on the right-hand side
by the leading-order diagrams (squared to give rates) for losing or creating
a particle of momentum $\p$.  I've suppressed details of phase
space integrals, the density of particles to scatter off of, or
final state Bose enhancement factors (or Pauli blocking factors
for fermions).  If I put all that in, and use crossing symmetry
to equate the gain and loss diagrams, we get the Boltzmann equation:
\begin {subequations}
\label {eq:boltz}
\begin {equation}
   (\partial_t + \v \cdot \grad) f = -C[f] ,
\label {eq:boltz1}
\end {equation}
where the ``collision term'' is
\begin {align}
  C[f] & = \int_{\k\p'\k'}
  \biggl|
  \raisebox{-0.2cm}{\psfig{file=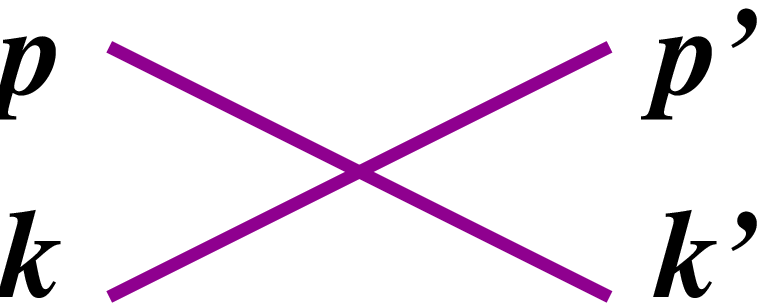,width=1.5cm}}
  \biggr|^2
  \Bigl\{
    f_\p f_\k (1\pm f_{\p'}) (1\pm f_{\k'})
    - f_{\p'} f_{\k'} (1\pm f_\p) (1 \pm f_\k)
  \Bigr\}
\nonumber\\
  & \quad + \mbox{(all other scattering processes)} .
\label {eq:boltzC}
\end {align}
\end {subequations}
The left-hand side of (\ref{eq:boltz1}) is called the convective
derivative of $f$.  If there were no collisions, it would simply
reflect the fact that $f$ at a particular point $\x$ can change
with time simply because the particles' velocities move them
to another point $\x + \v \Delta t$.


\subsection {Shear viscosity $\eta$ in scalar theory}

As an example of how things work, I'll start with the example of shear
viscosity in scalar $\phi^4$ theory.  I'll start by describing what
shear viscosity is.  One way to think of it is to imagine a moving
stream (infinitely long but finite radius $R$) surrounded by a stationary
ocean.  If we wait, the stream will broaden and slow down, eventually
dissipating its momentum across the bulk of the ocean.
The shear viscosity $\eta$ characterizes the rate of this dissipation,
which is proportional%
\footnote{
  More specifically, $\eta$ is defined so the rate
  $\sim \eta/R^2(\epsilon+P)$ where $\epsilon$ and $P$ are the energy
  density and pressure.  In the non-relativistic limit,
  $\epsilon+P$ becomes mass density.
}
to $\eta/R^2$.

\begin{figure}[th]
\centerline{\psfig{file=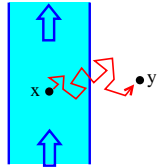,width=2cm}}
\caption{%
  \label {fig:shear}
  A particle random walking out of a stream of moving fluid.
}
\end{figure}

How does the dissipation happen?  At finite
temperature, particles (of water, air, quark-gluon plasma, whatever)
are bouncing around in all directions.  In the stream, the
{\it average}\/ value of the component of momentum in the stream's
direction is non-zero.  If we wait long enough, the particles in the
stream will randomly bounce out of the stream, spreading this net
momentum into the bulk of the medium.
Because of collisions, the particles move transversely
in a drunken, random walk, as in Fig.\ \ref{fig:shear}.
The shorter the steps of that walk (the shorter
the mean free path), the harder it will be to get out.
Note also that the more momentum each particle carries, the more it
will transfer when it does get out.
Roughly speaking,
\begin {equation}
   \mbox{rate} ~~\propto~~ \eta ~~\sim~~
   (\mbox{mean free path}) \times (\mbox{energy-momentum density}) .
\label {eq:shear0}
\end {equation}

\begin{figure}[th]
\centerline{\psfig{file=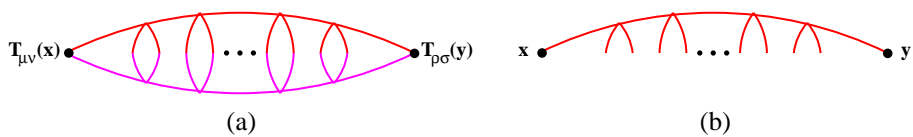,width=12cm}}
\caption{%
  \label {fig:shear_scalar}
  (a) A set of Feynman diagrams contributing to leading-order shear
  viscosity, and (b) a cut of those diagrams.
}
\end{figure}

How could we measure all of this formally?  Energy and momentum
are measured by the stress-energy tensor $T_{\mu\nu}$.
Imagine applying the operator $T_{\mu\nu}(x)$ to the equilibrium
state in order to create a disturbance at a space-time point
$x$ in the equilibrium fluid.  Now we can later measure
$T_{\mu\nu}(y)$ at other points $y$ to see how the disturbance
is spreading out with time.  So, it turns out to be possible to
relate viscosity to knowledge of the low-momentum behavior of
the (retarded) stress-energy correlator.%
\footnote{
  This and similar relations are known as Kubo relations.
}
We can then imagine computing such correlators using diagrammatic
methods of (finite temperature) quantum field theory.
Fig.\ \ref{fig:shear_scalar}a shows an
infinite class of diagrams which contribute
to the leading-order results for shear viscosity in $\phi^4$ theory.
Why are there so many interaction vertices?  One can get
a rough idea by considering chopping the diagram in half,
with the upper half shown in Fig.\ \ref{fig:shear_scalar}b.
This looks just like the random walk depicted in Fig.\
\ref{fig:shear} of a particle from $x$ to $y$ undergoing many
collisions with other particles on its way out.
If we're interested in long-wavelength, long-time behavior,
which is what viscosity (and hydrodynamics in general)
characterizes, then the number of collisions involved
is unavoidably large.

Fig.\ \ref{fig:shear_scalar}b is an amplitude, and we must square it
to get a rate.  One can crudely think of the bottom half of
Fig.\ \ref{fig:shear_scalar}a as the conjugate amplitude.

\begin{figure}[th]
\centerline{\psfig{file=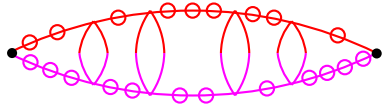,width=5cm}}
\caption{%
  \label {fig:shear_scalar2}
  The leading-order diagrams for shear viscosity at high temperature
  in $\phi^4$ theory.
}
\end{figure}

There's a slight variation needed to get {\it all}\/ the diagrams that
contribute to the leading-order shear viscosity, and I just want to give
you a flavor of it without dwelling on details.  In the multiple
scattering depicted by Fig.\ \ref{fig:shear_scalar}b, we should think of
the intermediate lines as being on-shell: the particle is on-shell in
between successive independent collisions.  But that means the
propagators corresponding to those lines are infinite.  This is a
standard problem in quantum mechanics whenever you produce a state that
does not live forever.  A standard example in zero-temperature quantum
field theory is the production of an unstable particle.  There the physics
which saves the day and avoids the on-shell divergence is the finite
width of the unstable state, which replaces
\begin {equation}
   \frac{1}{P^2-M^2} \to \frac{1}{P^2-M^2-i M \Gamma}
\label {eq:BW}
\end {equation}
and generates a
Breit-Wigner form.  At finite temperature, even stable particles
have a width, because collisions with other particles
will eventually knock them out of the momentum state they are in.
That width is related to the imaginary part of the
self-energy, which first arises in two-loop
self-energy diagrams (the imaginary part corresponding
to the rate for leading-order $2{\to}2$ scattering).  One
can effect the substitution analogous to (\ref{eq:BW})
by resumming these two-loop self-energies into the particle's propagator,
and so the relevant diagrams for the leading-order shear
viscosity are actually those shown in Fig.\ \ref{fig:shear_scalar2}.

In the context of $\phi^4$ theory, this diagrammatics was worked
out by Jeon in 1995.\cite{Jeon}
He showed that the leading-order contribution
of adding up the diagrams of Fig.\ \ref{fig:shear_scalar2} was
completely equivalent to finding the viscosity by linearizing
the Boltzmann equation (\ref{eq:boltz}) in small fluctuations about
equilibrium, and then solving those linearized equations.
The Boltzmann equation approach is a lot easier than analyzing
diagrams.  In either case, the result is $\eta \simeq 2860 \, T^3/\lambda^2$.


\subsection {Validity of the Boltzmann equation}

Why does the Boltzmann equation work?  There are two important
conditions for its validity.

The first condition is that the de Broglie wavelengths of particles
must be small compared to the mean free path between collisions
for the relevant particles:
\begin {equation}
   \lambda(\mbox{de Broglie}) \ll \mbox{mean free path} .
\label{eq:condition1}
\end {equation}
This allows one to treat the particles as {\it classical}\/
particles between collisions (allowing $\x$ and $\p$ to be
treated as independent, classical variables in the Boltzmann equation).
For scalar theory, the typical particle momentum is order $T$, and
so the de Broglie wavelengths are of order $1/T$.
The leading-order $2{\to}2$ scattering
amplitude is proportional to $\lambda$, rates are therefore
proportional to $\lambda^2$, and so, by dimensional analysis,
the mean free path is of order $1/\lambda^2 T$.
The condition (\ref{eq:condition1}) is thus satisfied in the
weak-coupling limit.

The second requirement is that the quantum-mechanical duration of
individual scattering events should be small compared to the mean
free time between collisions:
\begin {equation}
   \mbox{scattering duration} \ll \mbox{mean free time} .
\label{eq:condition2}
\end {equation}
Otherwise, there would be quantum interference between successive
scatterings, and we could not treat the scatterings as independent.
For scalar theory, the leading-order scattering diagram shown in
(\ref{eq:boltzC}) is a point interaction, which has zero time
duration, and so this is not an issue.

The difficulty, as we'll eventually see, is that the second condition
{\it fails}\/ for gauge theories!  That means we'll have to
work a little bit harder.


\subsection {Boltzmann for Gauge Theories}

Why can't we do what we did for scalar theory, putting the
leading-order scattering processes into the collision term?
Why isn't the
Boltzmann equation for gauge theory schematically of the form
\begin {equation}
   (\partial_t + \v \cdot \grad) f =
  \Biggl| ~
  \raisebox{-0.4cm}{\psfig{file=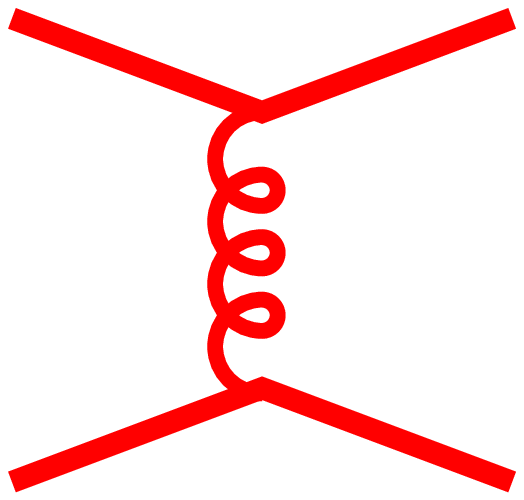,width=1cm}}
  ~ \Biggr|^2
  + \mbox{(other leading-order $2{\to}2$ diagrams)} ?
\label {eq:gboltz1}
\end {equation}
Before we can understand the problem, we will have to discuss
some basic scales associated with the process of Coulomb scattering
depicted in (\ref{eq:gboltz1}).

By the way, in general I will use straight lines in diagrams to
represent any $p \sim T$ particle.  So the diagram shown explicitly
in (\ref{eq:gboltz1}) represents $gg$ and $gq$ processes as well
as $qq$ processes (and processes involving anti-quarks).


\subsubsection {Coulomb scattering basics}

In vacuum, Coulomb scattering has an infinite cross-section,
arising from an infrared divergence associated with the
long-range nature of the Coulomb force.  The diagram in
(\ref{eq:gboltz1}) is proportional to $g^2/Q^2$,
where $Q$ is the exchanged 4-momentum and $1/Q^2$ comes from
the corresponding propagator.  After squaring the amplitude
and integrating over the final state phase space, one finds that
the cross-section is of order
\begin {equation}
  \sigma \sim \int d(Q^2) \left| \frac{g^2}{Q^2} \right|^2 ,
\label {eq:sigma1}
\end {equation}
which is infrared divergent.
In a plasma, however, the momentum exchange is cut off in the
infrared by Debye screening at $Q \sim \mD$.
So (with a caveat discussed later),
the cross-section (\ref{eq:sigma1}) is
\begin {equation}
  \sigma \sim \frac{g^4}{\mD^2} \sim \frac{g^4}{(g T)^2}
  \sim \frac{g^2}{T^2} \,.
\label {eq:sigma2}
\end {equation}

Here's an intuitive way to the same result which I find provides
a quick way to remember how things work.
Consider a particle moving through the medium, and let the term
``scatterer'' refer to another particle that it scatters from.
The Coulomb field of the scatterer extends out to the Debye screening
length $\xiD$.  The cross-sectional area of that region of space
is $\sim \xiD^2$.  If interactions were not weak, then the
scattering cross-section would also be $\sim \xiD^2$ (like
scattering off of a hard sphere).
But, in weak coupling,
the leading-order diagram of (\ref{eq:gboltz1}) has two explicit
powers of $g$, which gives $g^4$ in the rate.
So this modifies $\xiD^2$ to
\begin {equation}
  \sigma \sim g^4 \xiD^2 ,
\end {equation}
which is the same as (\ref{eq:sigma2}).

With the cross-section in hand, we can find the mean-free time $\tau$
between collisions.  The rate $\tau^{-1}$
for a given particle to collide is proportional to
the cross-section $\sigma$, the density $n \sim T^3$ of scatterers,
and the relative velocity $v \sim 1$ that the particle moves
through those scatterers.  So
\begin {equation}
  \tau \sim (n \sigma v)^{-1} \sim (T^3 \sigma)^{-1} \,.
\end {equation}
Using (\ref{eq:sigma2}),
\begin {equation}
  \tau \sim \frac{1}{g^2 T} \,.
\label {eq:mft}
\end {equation}

The quantum duration of this scattering event is of order
$1/Q \sim 1/\mD \sim 1/gT$.
For weak coupling, this duration is indeed small compared to
the mean free time (\ref{eq:mft}).  So far, there is
no problem with condition (\ref{eq:condition2}) for
the validity of the Boltzmann equation.

Typical Coulomb scatterings do not significantly randomize a particle's
direction and so individually do not create the random walk
depicted in Fig.\ \ref{fig:shear}.  To see this, let's estimate the
angle of deflection for a typical Coulomb scattering.  The momentum
transfer of order $Q \sim \mD$ will not significantly change the
total momentum $p\sim T$ of a typical plasma particle, but it
will give a transverse kick to that momentum of order $p_\perp \sim \mD$.
That leads to a change in direction of order
\begin {equation}
  \Delta\theta \sim \frac{p_\perp}{p} \sim \frac{\mD}{T} \sim g ,
\end {equation}
which is small in the weak-coupling limit.

\begin{figure}[th]
\centerline{\psfig{file=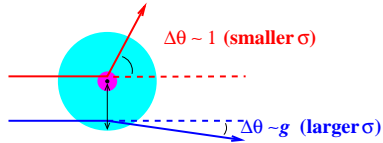,width=5cm}}
\caption{%
  \label {fig:coulomb}
  A picture of typical small-angle, large-impact parameter Coulomb
  scattering vs.\ rarer large-angle, small-impact parameter scattering.
}
\end{figure}

Much more rarely, individual Coulomb collisions will have large momentum
transfers, or equivalently small impact parameters, producing
large angle deflections $\Delta\theta \sim 1$ and $p_\perp \sim p$.
This comes from
the range of integration $Q^2 \sim p^2$ in (\ref{eq:sigma1}),
corresponding to $\sigma \sim g^4/p^2 \sim g^4/T^2$.
Equivalently, it comes from impact parameters $b$ of order $1/p$,
leading to $\sigma \sim g^4 b^2 \sim g^4/p^2$.  Computing
$\tau \sim (n \sigma v)^{-1}$, we can summarize this and the
previous result as
\begin {equation}
  \tau \sim
  \begin {cases}
    1/g^4 T , & \mbox{for}~\Delta\theta \sim 1; \\
    1/g^2 T , & \mbox{for}~\Delta\theta \sim g; \\
  \end {cases}
\end {equation}
A picture is shown in Fig.\ \ref{fig:coulomb}.

\begin{figure}[th]
\centerline{\psfig{file=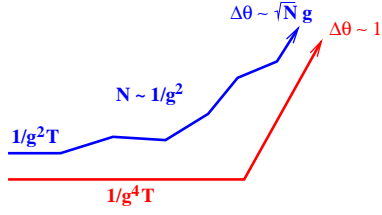,width=5cm}}
\caption{%
  \label {fig:small_v_large}
  Different ways to get a large-angle scattering in time
  $\sim 1/g^4 T$.
}
\end{figure}

So which sort of Coulomb collisions matter for characterizing the
random walk that determines shear viscosity?
The answer is that they both do.
The mean free time $\tau \sim 1/g^4 T$ for a single large-angle
collision is the same as the time $N/g^2 T$ for a large
number $N \sim 1/g^2$ of small-angle collisions (each $\delta\theta \sim g$).
The small-angle collisions will cause a random walk in the deflection
angle.  So the total deflection grows as $\sqrt{N}$, giving
\begin {equation}
  \Delta\theta \sim g \sqrt{N} \sim 1 .
\end {equation}
So, as summarized in Fig.\ \ref{fig:small_v_large},
a typical plasma particle's direction gets randomized in a
time of order $\tau \sim 1/g^4 T$, whether it be by a single
large-angle scattering or a sequence of many small-angle
scatterings.%
\footnote{
  Throughout this discussion, I have talked about Coulomb scattering
  and the fact that Coulomb fields are Debye screened.  But why
  can't $2{\to}2$ scattering occur via magnetic forces, which
  can operate over scales as large as the magnetic confinement
  scale $\xi_{\rm M} \sim 1/g^2 T$?  One might guess the
  cross-section for such super-large impact parameter scatterings
  to be $\sigma \sim g^4 \xi_{\rm M}^2 \sim 1/T^2$, which is much
  larger than the other cross-sections discussed, and the
  deflection to be $\Delta\theta \sim p_\perp/p \sim \xi_{\rm M}/T \sim g^2$.
  But it turns out that the cross-section estimate is wrong.  It's
  true that {\it static}\/ magnetic fields are not screened at the
  scale $1/gT$, but changing magnetic fields can be.  This is
  Lenz's Law from freshman physics: conductors resist changes
  to magnetic fields.  (The changing B fields produce E fields, which
  are in turn screened.)  This suppression produces
  $\sigma \sim (g^2/T^2) \ln(\mD/g^2 T)$, and the result is that
  $\Delta\theta \sim g^2$ collisions are no more frequent than
  $\Delta\theta \sim g$ collisions and so are not important in
  terms of randomizing particle directions.
}

\subsubsection {Showering}

\begin{figure}[th]
\centerline{\psfig{file=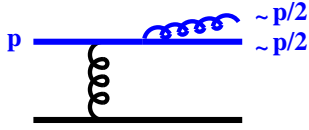,width=4cm}}
\caption{%
  \label {fig:brem}
  Showering of a particle by hard, nearly-collinear bremsstrahlung
  during a small-angle $2\to2$ scattering.
}
\end{figure}

We have seen that the relevant time scale for the random walk in
particle velocities is
\begin {equation}
  \tau_{\rm random} \sim \frac{1}{g^4 T} \,.
\label {eq:tmfp}
\end {equation}
This is known as the ``transport'' mean free path.
Does anything {\it else}\/ happen on this time scale that
might be relevant?  The answer is showering, by which I mean
hard, nearly collinear bremsstrahlung from small-angle
collisions, such as shown in Fig.\ \ref{fig:brem}.
Emitting the extra bremsstrahlung gluon costs an extra
factor of $g^2$ compared to the small-angle $2{\to}2$
scattering rate $\sigma_{2{\to}2} \sim 1/g^2 T$ of the previous
section.  That means the rate is $g^2$ times smaller, and so
the mean free path for showering is $1/g^2$ times bigger:
\begin {equation}
  \tau_{\rm shower} \sim \frac{1}{g^2} \, \tau_{\rm 2{\to}2}
  \sim \frac{1}{g^2} \, \frac{1}{g^2 T} \sim \frac{1}{g^4 T} \,.
\end {equation}
That is the same order as the transport mean free path (\ref{eq:tmfp}).

You might wonder why it's relevant whether or not the particle
splits into two.  The point of our discussion of shear viscosity is
to understand how quickly momentum can be transported out of the
stream of fluid in Fig.\ \ref{fig:shear}.  After a nearly-collinear
split like Fig.\ \ref{fig:brem}, the same total amount of momentum is
still flowing in roughly the same direction as it was before the
collision.  So how can this process impede momentum flow and
so affect the shear viscosity?  The answer is that the original
momentum $p$ is now carried by two particles with lesser momentum,
and it is easier for subsequent Coulomb scatterings to deflect
the directions of lower-momentum particles.  The deflection angle
$\Delta\theta \sim p_\perp/p \sim \mD/p$ of a
typical Coulomb scattering will be larger if $p$ is smaller.
So showering has this indirect effect on momentum transport, which
must be taken into account in a complete leading-order treatment.

We might hope to include this physics into our Boltzmann equation
as (schematically)
\begin {equation}
   (\partial_t + \v \cdot \grad) f =
  \Biggl| ~
  \raisebox{-0.4cm}{\psfig{file=tchannel_red.eps,width=1cm}}
  ~ \Biggr|^2
  + \Biggl| ~
  \raisebox{-0.4cm}{\psfig{file=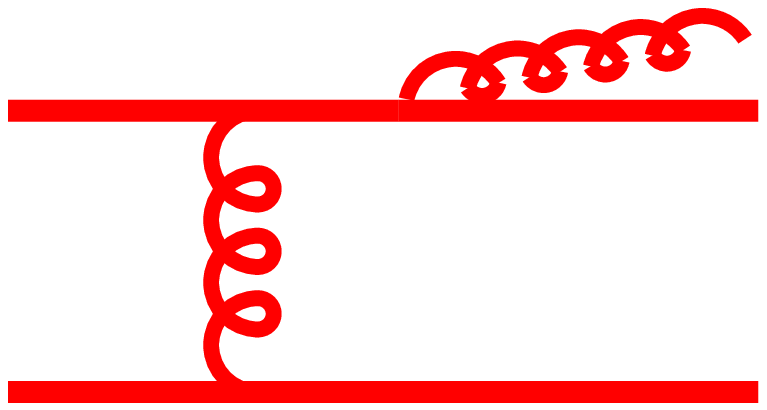,width=2cm}}
  ~ \Biggr|^2
  .
\label {eq:gboltz2}
\end {equation}
It turns out, however, that the internal straight line in Fig.\ \ref{fig:brem}
is only off-shell by a very tiny amount, and this has unfortunate
consequences for the quantum mechanical duration of the
bremsstrahlung process.
The splitting of one massless particle into two exactly collinear
massless particles is allowed by energy and momentum conservation,
which would allow the internal line to be exactly on shell.
Our particles are not massless: they have small thermal masses
of order $gT$, and so they require a small injection of momentum
to split.  It turns out that the internal straight line in
Fig.\ \ref{fig:brem} is off-shell by an amount
$\Delta E \sim m_\infty^2/p \sim g^2 T$ in energy.
This means that the scattering duration, known in this context as
the ``formation time'' of the bremsstrahlung gluon, is
\begin{equation}
  \mbox{formation time} \sim
  \frac{1}{\Delta E}
  \sim \frac{1}{g^2 T} .
\end {equation}
But this is parametrically the same as the mean free time
$\tau_{2{\to}2} \sim 1/g^2 T$ between small-angle scatterings!
We have included a process that fails the condition
(\ref{eq:condition2}) that scattering durations be small compared
to mean free times between collisions,
and so a Boltzmann treatment of the two processes in
(\ref{eq:gboltz2}) is not valid.%
\footnote{
  I should be a little careful to distinguish matters of principal from
  matters of practicality.  One can imagine that the formation
  time might be numerically small compared to $\tau_{\rm 2{\to}2}$
  even if not suppressed by a power of $g$.  If so, one might
  get a quite reasonable numerical approximation to the
  exact weak-coupling result by ignoring the problem.
  This turns out to be the case for shear viscosity.
  But I continue nonetheless to discuss what's needed for a
  complete leading-order calculation because it highlights
  physics that is important to other matters, such
  as the penetration of high energy jets through a thermal medium.
}


\subsubsection {The Landau-Pomeranchuk-Migdal effect}

To deal with this situation, we have to leap from the
physics of 1872 to the truly cutting-edge physics of... 1955.
We have been discussing QCD and quark-gluon plasmas, but the
same issues apply to QED and the electromagnetic showering of
very-high-energy cosmic rays in the atmosphere.
Consider a high-energy electron scattering off of the
Coulomb field of a nucleus and emitting a hard, bremsstrahlung
photon.
It was noted that, at sufficiently high energy, the
formation time would exceed the mean free path between collisions
in the atmosphere.
The result is a suppression of the rate for bremsstrahlung, because
the bremsstrahlung photon effectively
sees only one net deflection of the charged particle during its
formation time, rather than the several individual deflections
associated with each individual scattering.
This is known as the Landau-Pomeranchuk-Migdal (LPM)
effect.\cite{LPM}

\begin{figure}[th]
\centerline{\psfig{file=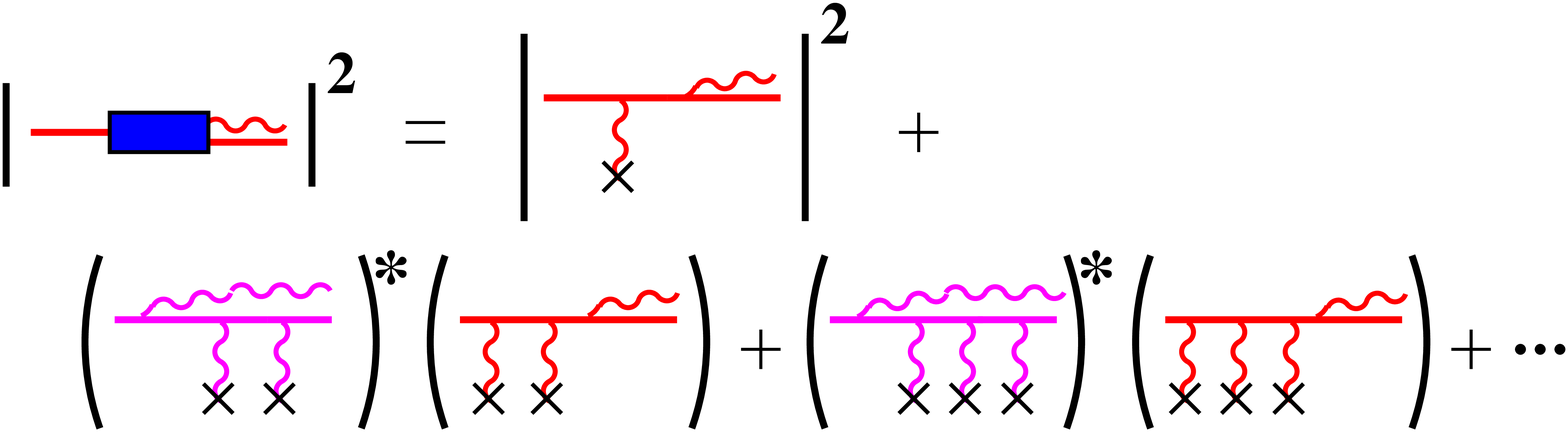,width=12cm}}
\caption{%
  \label {fig:lpm}
  The effective splitting rate in QED as a sum over possible
  interferences of photon emission before or after a certain
  number of scatterings.  Here, the crosses represent the
  sources for the scattering electromagnetic fields, such as
  fixed nuclei in the case of high-energy air showers.
}
\end{figure}

How can we proceed?
The long formation time means that photon production associated
with one collision could have interference effects with
photon production associated with a later collision.
In terms of diagrams, there could be interference terms
between photon production (i) just before and (ii) just after
some set of $N$ consecutive scatterings, as shown on the
right-hand side of Fig.\ \ref{fig:lpm}.  We want to sum up
all these possibilities into an effective rate for particle
splitting in the medium, depicted by the left-hand side of
Fig.\ \ref{fig:lpm}.

\begin{figure}[th]
\centerline{\psfig{file=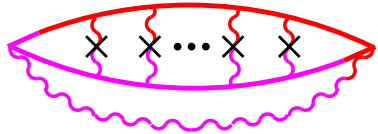,width=5cm}}
\caption{%
  \label {fig:lpm_graph}
  Another representation of the right-hand side of Fig.\ \ref{fig:lpm}.
}
\end{figure}

It turns out (without going into details) to be
useful to imagine
gluing together the diagrams for the amplitudes and
conjugate amplitudes to rewrite the right-hand side of Fig.\
\ref{fig:lpm} in the form of a sum of diagrams of the
form of Fig.\ \ref{fig:lpm_graph} (vaguely analogous to what
happened in Fig.\ \ref{fig:shear_scalar}).
Before I explain why it's useful, I'll first discuss the
generalization from QED to QCD.

\begin{figure}[th]
\centerline{\psfig{file=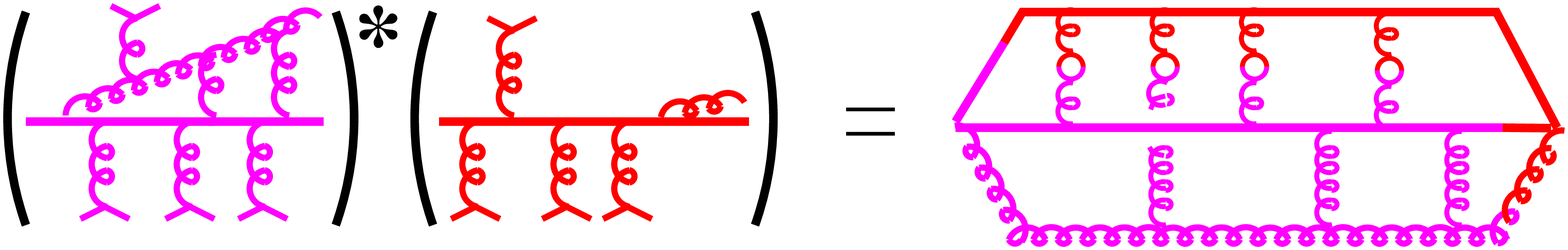,width=9cm}}
\caption{%
  \label {fig:glpm}
  A sample interference term for the LPM effect in QCD, analogous
  to terms in Figs.\ \ref{fig:lpm} and \ref{fig:lpm_graph}.
}
\end{figure}

The effective Boltzmann equation for QCD, which fixes the LPM problem
by using an effective splitting rate that sums up the interferences,
is schematically of the form\cite{bottom_up,effective}${}^{,}$%
\footnote{
  The last term in (\ref{eq:gboltz3} should be understood to
  include all $1 \leftrightarrow 2$ splittings, such as
  $q \leftrightarrow qg$, $g \leftrightarrow gg$, and
  $g \leftrightarrow q \bar q$.
}%
${}^,$%
\footnote{
  One can wonder if this equation double counts some small-angle
  collisions, because small-angle collisions appear both in the
  effective splitting rate and explicitly as the first diagram
  in (\ref{eq:gboltz3}).  The answer is yes but it's okay at
  leading-order, because only a parametrically small fraction
  $t_{\rm form}/t_{\rm random} \sim g^2$ of small-angle collisions
  take place while a bremsstrahlung gluon is forming.
}
\begin {equation}
   (\partial_t + \v \cdot \grad) f =
  \Biggl| ~
  \raisebox{-0.4cm}{\psfig{file=tchannel_red.eps,width=1cm}}
  ~ \Biggr|^2
  + \Biggl| ~
  \raisebox{-0.1cm}{\psfig{file=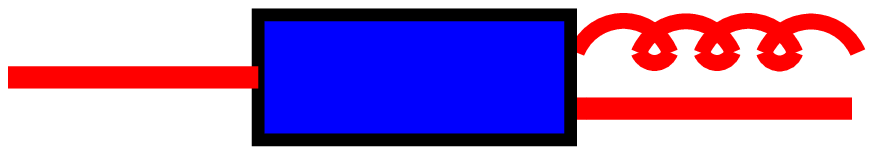,width=2cm}}
  ~ \Biggr|^2
  .
\label {eq:gboltz3}
\end {equation}
The difference of the last term with the QED case of Fig.\ \ref{fig:lpm}
is that the bremsstrahlung gluon carries color and so can also scatter
from the medium.  In addition, there can be final state interactions
between the two split particles.  Fig.\ \ref{fig:glpm} shows a
typical interference term, along with the diagrammatic representation
if you sew together the amplitude and conjugate amplitude,
analogous to Fig.\ \ref{fig:lpm_graph}.%
\footnote{
  For a discussion in the language used here, see Ref.\ \refcite{sansra}.
  For earlier work, see Ref.\ \refcite{lpm_other}.
}

\begin{figure}[th]
\centerline{\psfig{file=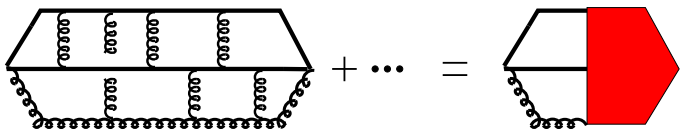,width=9cm}}
\caption{%
  \label {fig:glpm_graph}
  Rewriting the sum of QCD LPM graphs in terms of an unknown blob.
}
\end{figure}

\begin{figure}[th]
\centerline{\psfig{file=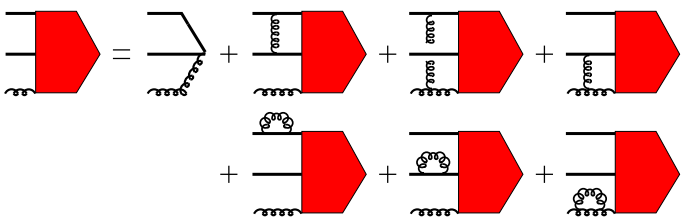,width=9cm}}
\caption{%
  \label {fig:glpm_int}
  An integral equation for the blob.
}
\end{figure}

I haven't given you any details about how exactly you are supposed to
evaluate a diagram like the right-hand side of Fig.\ \ref{fig:glpm}.
And I have to sum up an infinite class of similar diagrams, which
sounds difficult.  But, having written the diagrams in this form, I can
now explain to you how the problem of doing this entire diagram
sum can be reduced to the problem of solving an integral equation
(which in practice can be done numerically\cite{shear}).
In Fig.\ \ref{fig:glpm_graph}, the left-hand side represents the infinite
sum of diagrams needed for the LPM effect.
On the right-hand side, the blob represents the sum of those
same diagrams if I simply cut out their left vertex.
This blob satisfies an integral equation shown by the first line
in Fig.\ \ref{fig:glpm_int}, which you can see by thinking of all but
the first line term on the right-hand side as a ``perturbation''
and then solving the equation iteratively.
Fig. \ref{fig:glpm_int} is an integral equation because there is
one loop integral, in terms of the unknown blob,
on all but the first term on the right-hand side.

You'll see, however, that in Fig.\ \ref{fig:glpm_int}
I have also included a
second line which sums in some corresponding self-energy corrections.
That's because I've hidden something from you in the interest of
simplifying my presentation so far.  For the same reason that
I needed to include the imaginary part of self-energy diagrams in
Fig.\ \ref{fig:shear_scalar2}, I need to do it here.%
\footnote{
  When a particle propagates over a time as long as its mean free
  path, I can't ignore the effect of its width on the propagator.
}
Finally, I
should mention that all the lines shown explicitly in
Fig.\ \ref{fig:glpm_int}
are themselves resummed with $O(g^2 T^2)$ one-loop self-energies (which
give Debye screening and whose imaginary parts give the little circular
loops representing the scatterers on the right-hand side of Fig.\
\ref{fig:glpm}, which I am no longer showing explicitly).


\subsubsection {Result for shear viscosity}

I should quote you a result just to show that all of this
formalism can be successfully implemented.  We've previously
discussed (\ref{eq:shear0}) that shear viscosity is proportional to
the mean free time (\ref{eq:tmfp})
characteristic of a particle's random walk.
The rest is just factors of $T$ determined by dimensional analysis,
giving $\eta \sim T^4 \tau_{\rm random}$.  So we expect
$\eta \simeq \# T^3/g^4$, and we'd like to know the numerical
coefficient $\#$, analogous to the result quoted earlier for
scalar theory.

\begin{figure}[th]
\centerline{\psfig{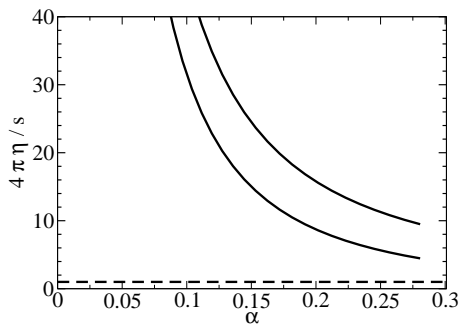}}
\caption{%
  \label {fig:shear_result}
  Leading-order
  results adapted from Arnold, Moore and Yaffe\protect\cite{shear} for
  shear viscosity in 3-flavor QCD, as a function of $\alpha_{\rm s}$,
  in units of entropy density $s$ over $4\pi$.
  [The leading-order value of
  $s/4\pi$ is $\pi T^3 (8/45 + 7 N_{\rm f}/60)$.]
  The two solid lines show two different calculations that
  should agree at leading order in coupling but differ at higher
  order, giving an idea of some of the possible
  errors from higher-order effects.
  The dashed line shows a conjectured lower bound on
  $\eta/4\pi s$ for any relativistic system.\protect\cite{shearbound}
}
\end{figure}

However, it's slightly more complicated than that.
The reason is because of the multiple ways for a particle's
velocity to randomize, as was depicted in Fig.\ \ref{fig:small_v_large}.
Both large and small angle collisions were equally efficient.
So is a sequence of intermediate angle collisions.  As a result,
the rate is enhanced by a logarithm, known as the ``Coulomb log,''%
\footnote{
  For a textbook discussion, see Sec.\ 41 of Ref.\ \refcite{LL10}.
}
counting all the decades of possibility between ``large angles''
($\delta\theta \sim 1$) and ``small angles'' ($\delta\theta \sim g$),
so that
\begin {equation}
  \tau_{\rm random} \sim \frac{1}{g^4 T \ln(1/g)}
  \qquad \mbox{and} \qquad
  \eta \sim \frac{\# T^3}{g^4 \ln(1/g)} \,.
\end {equation}
But logarithms are not particularly big, and there are corrections
to this formal limit order by order in $1/\ln(1/g)$.  So, if one
wants a result good to the leading {\it power}\/ of $g$, it has
the functional form
\begin {equation}
  \eta \simeq \frac{T^3}{g^4 \ln(1/g)} \, F\bigl(\ln(1/g)\bigr) ,
\end {equation}
and the answer requires giving a logarithmically-sensitive function
$F$ rather than a single numerical coefficient $\#$.
Just to prove to you that it can be done, Fig.\ \ref{fig:shear_result}
shows the leading-order result
of the ratio of shear viscosity $\eta$
to entropy density $s$ as a function of coupling.

\subsubsection {Summary}

To summarize, I've tried to show you that, with enough work, it is
possible to construct an effective Boltzmann equation that can be
used for {\it leading}-order calculations of hydrodynamic transport.
The methodology is far less systematic than for Euclidean field
theories.  In particular, going beyond leading order is challenging!

\begin{figure}[th]
\centerline{\psfig{file=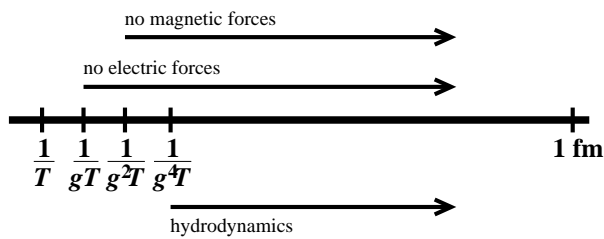,width=8cm}}
\caption{%
  A cartoon of distance scales relevant to
  high-temperature gauge theory.
  \label {fig:scales2}
}
\end{figure}

In looking at the real-time behavior, we also encountered a new
scale, $g^4 T$, beyond the ones summarized earlier in Sec.\ \ref{sec:scales1}:
\begin {itemlist}
\item
$g^2T$: rate for small-angle scattering.
\item
$g^4T$: rate for large-angle scattering.
\end {itemlist}
A cartoon of the hierarchy of distance scales is shown in
Fig.\ \ref{fig:scales2}.

\section{Hard Loops and the Vlasov Equations}

Looking at Fig.\ \ref{fig:scales2}, note that, because of magnetic
confinement, the long distance effective theory of a quark-gluon plasma
is hydrodynamics and not the {\it magneto}-hydrodynamics (MHD) relevant
to traditional electromagnetic plasmas.  So, in some ways, non-abelian
plasmas are a lot simpler than traditional electromagnetic plasmas,
because there are no magnetic forces over large distances, like those
responsible for the complicated structure of solar filaments.

Does that mean that there are no interesting collective effects
in quark-gluon plasmas analogous to the rich variety of phenomena
in traditional plasma physics?  Not quite.  There is a window
of distance scales $l$,
\begin {equation}
   \frac{1}{T} \ll l \ll \frac{1}{g^2 T} \,,
\label {eq:window}
\end {equation}
large
compared to the inter-particle separation but small compared to
the magnetic screening length, which is large enough for collective
effects yet small enough for magnetic and sometimes electric
effects.  An example we've already seen are low-momentum
longitudinal plasma waves, whose collective nature was depicted
in Fig.\ \ref{fig:longitudinal}, where the characteristic scale
was the plasma frequency $\omega_{\rm pl} \sim gT$.

\begin{figure}[th]
\centerline{\psfig{file=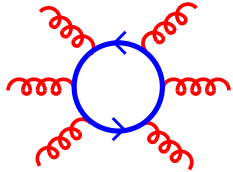,width=3cm}}
\caption{%
  A generic hard thermal loop.  The internal line has $p\sim T$
  and the external lines are soft ($k \ll T$).
  \label {fig:HTL}
}
\end{figure}

Our earlier discussion of plasma waves arose by considering
one loop self-energies like Fig.\ \ref{fig:gloop},
which were dominated by hard loop momenta $p \sim T$ and
which had a significant effect on the propagation of low-momentum
plasma waves.  Such a loop is called a ``hard thermal loop.''
For imaginary-time physics, the self-energy turns out to be
the end of the story as far as hard thermal loops are concerned.
For real-time physics, however, Braaten and Pisarksi\cite{HTL}
discovered
some years ago that other one-loop diagrams such as
Fig.\ \ref{fig:HTL} give important hard thermal loops,
creating significant interactions between low momentum
(e.g. $k \sim gT$) gluons.  They found that all such
interactions could be described by an effective Lagrangian
for low-momentum gluons:
\begin {equation}
   {\cal L}_{\rm eff} =
   - \frac{\mD^2}{2} \left\langle
     F_{\mu\nu}^a \frac{v^\mu v^\rho}{(v\cdot D)^2} \, F^{b\nu}_\rho
   \right\rangle_\v ,
\label {eq:HTL}
\end {equation}
where $v^\mu \equiv (1,\v)$ with $\v$ a unit vector,%
\footnote{
  Note: The notation $v^\mu$ is convenient, but $v^\mu$ does not
  transform as a 4-vector.
}
and the angle brackets denote averaging over the direction of $\v$.
This looks complicated and non-local because there is a covariant
derivative $D$ in the {\it denominator}.  However, it turns out
that there is simple, local physics that underlies this result, to
which I now turn.

\subsection {The Vlasov Equations}

For simplicity, let's start with a QED plasma.
Like in the picture of Fig.\ \ref{fig:longitudinal} of a
longitudinal plasma wave, we want a description of
(i) long wavelength (e.g. $\lambda \sim 1/gT$) fluctuations
in the densities of hard ($p \sim T$) particles, coupled to
(ii) soft ($k \sim gT$) gauge fields.
Also, at these scales, we can ignore random, incoherent,
two-particle collisions, because we are interested in physics
in the window (\ref{eq:window}), which is restricted to distances small
compared to the mean free path $\tau_{2{\to}2} \sim 1/g^2 T$
of such collisions.

Let's start by re-examining the Boltzmann equation.  Since we
can ignore collisions in this context, we start with
\begin {equation}
  \frac{df}{dt} = C = 0 .
\end {equation}
By the chain rule,
\begin {equation}
   \frac{\partial f}{\partial t}
   + \frac{\partial\x}{\partial t} \cdot \frac{\partial f}{\partial\x}
   + \frac{\partial\p}{\partial t} \cdot \frac{\partial f}{\partial\p}
   = 0 ,
\end {equation}
where (unlike earlier discussion) I allow for a common
force $\F = \partial\p/\partial t$ on the particles due
to the influence of the soft gauge fields that we want to incorporate.%
\footnote{
  These fields weren't relevant in our previous discussion of
  hydrodynamics because of magnetic confinement on the relevant
  distance scales.
}
Substituting in the usual form of the electromagnetic force,
\begin {subequations}
\label {eq:QEDvlasov}
\begin {equation}
   \bigl[\partial_t + \v \cdot \grad_\x
          + g(\E+\v\times\B) \cdot \grad_\p\bigr] f
   = 0 .
\end {equation}
We combine this with Maxwell's equations for the soft field, with
current given by the hard particles:
\begin {equation}
  \partial_\mu F^{\mu\nu} = j^\nu = g \int_\p v^\nu f .
\end {equation}
\end {subequations}
(I have suppressed here an implicit sum over particle types.)
The two equations (\ref{eq:QEDvlasov}) are known as the Vlasov
equations.

To investigate thermal systems,
linearize the distribution
of particles in small fluctuations $\delta f$
around thermal equilibrium $f_{\rm eq}$,
\begin {equation}
   f(\x,\p,t) = f_{\rm eq}(p) + \delta f(\x,\p,t) ,
\end {equation}
to get
\begin {subequations}
\label {eq:QEDlinear}
\begin {equation}
   (\partial_t + \v \cdot \grad_\x) \delta f
   = - g(\E+\v\times\B) \cdot \grad_\p f_{\rm eq} ,
\label {eq:QEDlinear_boltz}
\end {equation}
\begin {equation}
  \partial_\mu F^{\mu\nu} = g \int_\p v^\nu \delta f .
\end {equation}
\end {subequations}

For QCD, the net color of the hard particles is treated quantum
mechanically,%
\footnote{
  There are some versions that approximate the color as classical.
}
$f(\x,\p,t)$ becomes a color density matrix, and space-time
derivatives become covariant derivatives.  In the linearized
version (\ref{eq:QEDlinear}), $f_{\rm eq}$ is color neutral,
and $\delta f$ can be taken in the adjoint representation, giving%
\footnote{
  See Ref.\ \refcite{kinetic} for early work on kinetic theory and
  Ref.\ \refcite{kineticHTL} for its application to HTL's.
}
\begin {subequations}
\label {eq:QCDlinear}
\begin {equation}
   (D_t + \v \cdot \D_\x) \delta f
   = - g(\E+\v\times\B) \cdot \grad_\p f_{\rm eq} ,
\label {eq:QCDlinear_boltz}
\end {equation}
\begin {equation}
  D_\mu F^{\mu\nu} = g \int_\p v^\nu \delta f .
\label {eq:QCDlinear_Maxwell}
\end {equation}
\end {subequations}
(I have suppressed here some numerical group factors.)
These are local equations.
but if we formally solve (\ref{eq:QCDlinear_boltz}) for $\delta f$
and plug into (\ref{eq:QCDlinear_Maxwell}), we get a non-local
equation for the evolution of the soft fields $A$:%
\footnote{
  I should mention that the $\v\times\B \cdot \grad_\p f_{\rm eq}$
  term in (\ref{eq:QCDlinear_boltz}) and (\ref{eq:HTL2}) vanishes,
  because an equilibrium distribution
  $f_{\rm eq}(\p)$ (Bose or Fermi) cares only about $|\p|$, so
  $\grad_\p f_{\rm eq}$ is in the direction of $\hat\p = \hat\v$ and
  so vanishes when dotted into $\v\times\B$.
  I have kept the term because in non-equilibrium applications (such as
  for the plasma instabilities discussed later), one wants to
  expand around non-isotropic distributions $f_0(\p)$ rather than
  $f_{\rm eq}(\p)$, and in that case the $\v\times\B$ terms will not vanish.
}
\begin {equation}
  D_\mu F^{\mu\nu}
  = - g^2 \int_\p v^\nu (D_t + \v\cdot \D)^{-1} (\E+\v\times\B)
    \cdot \grad_\p f_{\rm eq} .
\label {eq:HTL2}
\end {equation}
This turns out to be equivalent to the equation of motion that you
get from the HTL Lagrangian (\ref{eq:HTL}).
The moral is that the physics is local, described by
(\ref{eq:QCDlinear}), provided you write it in terms of the
full set of long-distance degrees of freedom in the theory:
the soft fields $A$ and the long-wavelength fluctuations $\delta f$
in particle distributions.


\subsection {The self-energy revisited}

As an example, I'll outline how to use the Vlasov equation to derive the
soft gluon self-energy.  To do so, it's sufficient to treat 
the gluons perturbatively, so that the soft field
$A$ is a small quantity like $\delta f$.
Then the linearized QCD Vlasov equations (\ref{eq:QCDlinear})
reduce in form to the QED case (\ref{eq:QEDlinear}).
Fourier transform the Boltzmann equation (\ref{eq:QEDlinear_boltz}) ,
\begin {equation}
   i (-\omega + \v \cdot \k) \delta f
   = - g(\E+\v\times\B) \cdot \grad_\p f_{\rm eq} ,
\end {equation}
solve for $\delta f$, and plug into the Maxwell equation,
\begin {equation}
   \partial_\mu F^{\mu\nu} = g \int_\p v^\nu \delta f
   = -i g^2 \int_\p \frac{v^\nu (E+\v\times\B)}{-\omega+\v\cdot\k}
     \cdot \grad_\p f_{\rm eq} .
\end {equation}
Write this as $\partial_\mu F^{\mu\nu} = \Pi^{\nu\rho} A_\rho$
and extract $\Pi$.  If you do this, you get the standard
result for the soft gluon self-energy $\Pi(\omega,\k)$
that you would get if you instead did the one-loop diagrammatic
calculation discussed in Sec.\ \ref{sec:gmass}.
Kinetic theory (here in the form of the Vlasov equations) reproduces
the same result one obtains from thermal loop calculations in the
underlying quantum field theory.


\section{Far from Equilibrium: Plasma Instabilities}

I now want to investigate thermalization in cases that start far from
equilibrium.  To motivate the sort of question I want
to ask, I'm going to start with a cartoonish description of a heavy
ion collision.


\subsection {QGP hydrodynamics and elliptic flow}

\begin{figure}[th]
\centerline{\psfig{file=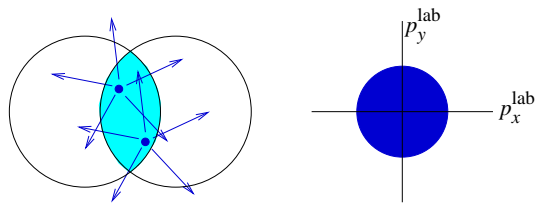,width=7cm}}
\caption{%
  A heavy ion collision if it were a superposition of
  independent nucleon-nucleon collisions.
  \label {fig:collide1}
}
\end{figure}

The left-hand side of Fig.\ \ref{fig:collide1} shows a picture
of a heavy ion collision, looking down the beam pipe.  The nuclei
are two Lorentz-contracted pancakes, represented by the large circles.
One nucleus is heading towards you, out of the page, and the other
nucleus is heading into the page.  In general, they won't collide
perfectly head on, and so there will be an almond shaped collision
region, as shown.  What if nothing interesting happened in
heavy ion collisions?  What if the result just looked like a bunch
of independent nucleon-nucleon collisions, the products of which
flew freely straight out to the detector?  The debris of each
nucleon-nucleon collision would (statistically) be isotropic in
the transverse plane of the paper.  So, if I looked at the
transverse momentum distribution of particles seen in the detector,
it would look rotationally invariant around the beam axis for
each heavy ion collision, as depicted by the right-hand side of
Fig.\ \ref{fig:collide1}.

\begin{figure}[th]
\centerline{\psfig{file=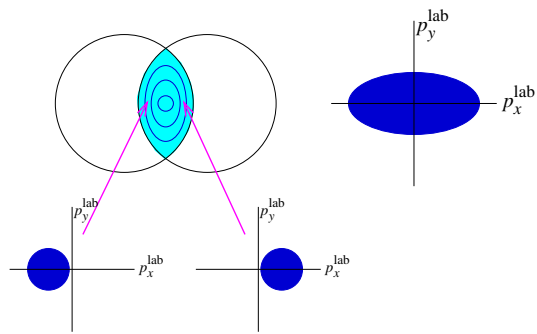,width=7cm}}
\caption{%
  Elliptic flow in a heavy ion collision.
  \label{fig:collide2}
}
\end{figure}

Now suppose that the products of the nucleon-nucleon collisions
can't fly freely out, because they collide with the products of
other nucleon-nucleon collisions on the way.  And suppose that
there are enough collisions that these particles, at least
briefly, come to approximate local thermal equilibrium
(e.g.\ a locally thermalized quark-gluon plasma).
Imagine a time slightly after the collision, where the two
nuclear pancakes have passed beyond each other, but the
almond region is filled up into an almond cross-sectioned cylinder
between the two.  Outside that ``cylinder'' is vacuum, with zero
pressure.  In the center is some central pressure, and there will
be a smooth gradient between the two, depicted by the contour lines
in the upper left-hand
figure in Fig.\ \ref{fig:collide2}.  As you can see from the figure,
the pressure gradient must be larger along the short diameter
of the almond.  Pressure gradients correspond to forces, and so
fluid elements along the short diameter will get accelerated
(to the right and left) more than elements along the long diameter
(accelerated up and down in the diagram).  As a result of this
anisotropic acceleration, the final products of the collision,
when they finally split up and fly out to the detector, will
(in a single heavy-ion collision) have
an anisotropic distribution in lab-frame {\it transverse}\/
momentum, as shown by
the upper right-hand figure in Fig.\ \ref{fig:collide2}.
The acceleration of fluid elements is an example of hydrodynamic
flow, and this particular manifestation is called ``elliptic flow.''

In order to keep things straight in the discussion that follows, it will
be important to distinguish between momentum of (i) all the particles in
the lab frame, and (ii) local sets of particles in a local fluid frame.
In the second case, I mean momenta as measured by an observer moving
with the fluid (plasma), who only looks at particles that are close to
her.  In the hydrodynamic picture, elliptic flow is caused by pressure,
which arises because of local thermalization.  But in a thermal medium,
momentum distributions are {\it isotropic}\/ in local fluid rest frames.
The lower part of Fig.\ \ref{fig:collide2} depicts two such isotropic
distributions, one for a fluid element moving to the left and one to the
right, boosted to the lab frame.  Though each distribution is isotropic
in its local fluid frame, their superposition is anisotropic in the lab
frame.  So, isotropization of momentum (via collisions) in local fluid
rest frames creates pressure and ultimately produces anisotropic momenta
(the signal of elliptic flow) in the {\it lab}\/ frame.

The sooner collisions develop pressure, the sooner hydrodynamic flow
can kick in, and the more significant will be elliptic flow.
(If it takes too long, then the almond will in the meantime
expand by free streaming into something bigger and more nearly
circular, and so the difference between the two transverse
directions will be less significant.)  Matching simulations of
ideal hydrodynamics to experimental data has led to a number of
successes,%
\footnote{
  See, for example, Ref.\ \refcite{hydro}.
}
but only if the system thermalizes very rapidly, on
a time scale less than 1 fm/$c$.
An important phenomenological
question is whether such fast thermalization is reasonable.
Can a quark-gluon plasma reach local thermal equilibrium in that
short a time?

So here's a question for theorists: What is the thermalization time
for a quark-gluon plasma?  That sounds hard, so let's start with
baby steps by asking a simpler question: What is it in the limit
of arbitrarily weak coupling?
As I mentioned in the introduction, even this problem is hard
enough that we don't even know how the answer scales with
coupling!
\begin {equation}
   t_{\rm eq} \sim \frac{g^{-??}}{\mbox{momentum scale}} \,.
\label {eq:tthermalize}
\end {equation}
I'll now explain what the difficulty is.


\subsection {Bottom-up thermalization}

In a wonderful 2000 paper, Baier, Mueller, Schiff and Son\cite{bottom_up}
analyzed the effect of individual particle collisions on
thermalization and determined the power ``??'' in
(\ref{eq:tthermalize}) to be 13/5.
There's something they missed in their analysis, which is the
difficulty I will get to shortly.  But first, let me give
an idea of the process of thermalization they found, which they
called ``bottom-up'' thermalization.

Imagine that the heavy-ion collision takes place at position $z=0$
along the beam axis.  This is the point where I make contact with
Francois Gelis's talk.\cite{gelis}
At high energies, it's expected that a very large (non-perturbatively
large) density of (low $x$) gluons are scattered out of the beam,
and that these gluons are what's left behind from the collision to
eventually form the quark-gluon plasma.  (This initial state is
called the color glass condensate and is described at early times
using classical field theory, but we won't need to concern ourselves
with those details here.)
In what follows, I will refer to these initial scattered gluons as
``hard'' particles (even though they are quite low momentum
compared to the initial nucleons that produced them)
because I will reserve the term ``soft'' for yet lower momentum
scales.

\begin{figure}[th]
\centerline{\psfig{file=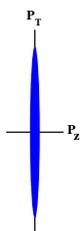,width=1cm}}
\caption{%
  An anisotropic local momentum distribution created by
  $p_z$ selection as the system expand in the beam direction
  after the collision.
  \label{fig:bmss0}
}
\end{figure}

As the system expands in the space between
the retreating, fragmented nuclei, it dilutes, and eventually the
gluon density is no longer non-perturbative, and we can think of
the system as a gas of individual gluons.
As a first, crude approximation,
let's ignore interactions and think of the gluons as free streaming.
The only gluons that will hang around in the neighborhood $z=0$
of the initial collision
will be those with a small $z$ component $v_z$ of velocity:
the gluons with significant $v_z$ move to a different $z$.
So, after a while, the momentum distribution of those gluons
in the neighborhood of $z=0$ starts to look very anisotropic,
as shown in Fig.\ \ref{fig:bmss0}.%
\footnote{
  Note that this is anisotropy in a local rest frame (and also is
  isotropic in the transverse plane) and has nothing
  to do with the anisotropy of elliptic flow.  Quite the opposite:
  this local anisotropy represents, so far, a {\it failure}\/ to thermalize.
}
In fact, the system looks similar (up to a boost) at other values
of $z$.  The only particles to get to the neighborhood of $z$
at time $t$ will be those with $v_z \simeq z/t$.  If I boost
by this amount to go to the ``local fluid rest frame,''
I'll again get the situation of Fig.\ \ref{fig:bmss0}.

\begin{figure}[th]
\centerline{\psfig{file=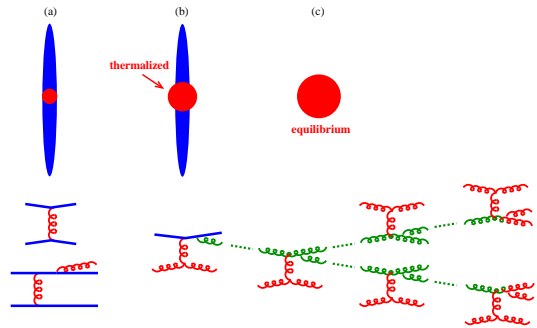,width=7cm}}
\caption{%
  Bottom-up thermalization.
  \label{fig:bottom_up}
}
\end{figure}

The process of bottom-up thermalization is shown in
Fig.\ \ref{fig:bottom_up}.  Diagram (a) depicts a momentum
distribution similar to Fig.\ \ref{fig:bmss0}.
However, the particles are not free streaming: they interact
via small angle collisions such as shown below diagram (a).
Such collisions broaden $p_z$ slightly, competing against the
effect of expansion to shrink the local $p_z$ distribution.
So diagram (a) is broader in $p_z$ than Fig.\ \ref{fig:bmss0}.
Secondly, such collisions are occasionally accompanied by
bremsstrahlung gluons, of which soft gluons are the most common.
The circular area inside the momentum distribution of diagram (a)
depicts these bremsstrahlung gluons.  It's easier to deflect
a lower momentum particle than a higher momentum particle, so
these softer gluons can fairly easily thermalize via collisions.
(It's called bottom-up thermalization because the lowest momentum
particles thermalize first.)  Eventually, enough of these
soft, bremsstrahlung gluons get produced, as indicated in
diagram (b), that the hard particles
favor scattering off of them rather than other hard particles.
And eventually, there are so many soft particles to scatter from,
that the relatively rare process of significant energy loss during
bremsstrahlung becomes important.  At that point, the energy
stored in the hard particles rapidly cascades to lower momentum
particles, as depicted below diagram (b), leaving us with the
final thermalized plasma in diagram (c).

So what's the problem?  The problem is that plasma physics is
complicated.  There can be collective effects that are not
captured by the independent, random particle collisions considered
in the original bottom-up scenario.  And one class of such effects
is plasma instabilities.


\subsection {The Weibel (filamentation) instability}

\begin{figure}[th]
\centerline{\psfig{file=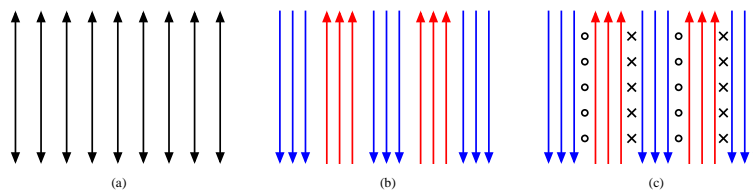,width=10cm}}
\caption{%
  The Weibel instability
  \label{fig:weibel}
}
\end{figure}

The instability I will describe arises in non-equilibrium
situations where particle momenta are anisotropic in local
fluid frames.  As an extreme example, think of ordinary
electromagnetism and
consider Fig.\ \ref{fig:weibel}a, which depicts two homogeneous
inter-penetrating currents of charged particles: one set of
particles heading up the page, and the other down.
Now, just to simplify my explanation, pretend that these
particles were running through wires.  Each line in
Fig.\ \ref{fig:weibel}a represent two wires on top of each other,
one with a current up the page and one with a current down.
Now recall from freshman physics (or use your right hand)
that parallel currents attract and opposite currents repel.
So the wires are unstable to clumping, as shown in
Fig.\ \ref{fig:weibel}b.  The up-going wires want to get near
each other and away from the down-going wires; the down-going
wires want the same.
Associated with these filaments are magnetic fields (which are
causing the attraction and repulsion of the wires),
shown coming in and out of the paper by the crosses and
circles in Fig.\ \ref{fig:weibel}c.
This instability is called the Weibel
or filamentation instability and was discovered theoretically
in 1959.\cite{Weibel}

The instability growth rate turns out to be much faster
than the rate of individual particle collisions, and instabilities
create relatively large soft magnetic fields.
The hard particles of the bottom-up scenario depicted in
Fig.\ \ref{fig:bottom_up} can scatter from these fields rather
than from each other or soft particles created by bremsstrahlung.
Unfortunately, all the details have not yet been worked out.

I would love to tell you more about Weibel instabilities,\cite{mrow_alm}
and how their fate in QCD plasmas differs from traditional electromagnetic
plasmas,\cite{linear}
and why this is important for understanding
thermalization.\cite{bodeker_extreme1}
I'd like to be able to tell you how the development of instabilities
can be simulated by simulating the QCD Vlasov equations
(\ref{eq:QCDlinear}).\cite{linear,linearB}
And I'd like to explain what's left
that needs to be done to complete the picture of thermalization
at weak coupling.  Unfortunately, I have run out of the pages
allocated to writing up these lectures.


\section*{Acknowledgements}

I thank the organizers of X Hadron Physics for inviting me to give
these lectures.  I am also indebted to my long-time collaborators,
Guy Moore and Larry Yaffe.
This work was supported, in part, by the U.S. Department
of Energy under Grant No.~DE-FG02-97ER41027.

\end{document}